\documentclass[longauth]{aaEC}  

\usepackage{graphicx}
\usepackage{euclid} 
\usepackage{txfonts}
\usepackage{booktabs}
\usepackage{comment}
\usepackage[table]{xcolor}
\usepackage{hyperref}
\usepackage[switch]{lineno}
\titlerunning{\Euclid Q1 Bars}
\begin{document}
\title{Euclid Quick Data Release (Q1)}\subtitle{A first look at the fraction of bars in massive galaxies at $z<1$\thanks{The authors of this paper wish to express their sincere gratitude to the
late Dr Peter Erwin, who passed away unexpectedly at the end of January
2025. Peter was an expert in the properties of barred galaxies, and his
thoughtful papers on the subject will form a lasting legacy. A member of
the Euclid Collaboration, Peter helped shape the current paper by
communicating intensively with the first author, and we wholeheartedly
acknowledge his contributions.}}
  
\newcommand{\orcid}[1]{} 

\author{Euclid Collaboration: M.~Huertas-Company\orcid{0000-0002-1416-8483}\thanks{\email{mhuertas@iac.es}}\inst{\ref{aff1},\ref{aff2},\ref{aff3},\ref{aff4}}
\and M.~Walmsley\orcid{0000-0002-6408-4181}\inst{\ref{aff5},\ref{aff6}}
\and M.~Siudek\orcid{0000-0002-2949-2155}\inst{\ref{aff2},\ref{aff7}}
\and P.~Iglesias-Navarro\orcid{0009-0009-8959-2404}\inst{\ref{aff1},\ref{aff8}}
\and J.~H.~Knapen\orcid{0000-0003-1643-0024}\inst{\ref{aff1},\ref{aff8}}
\and S.~Serjeant\orcid{0000-0002-0517-7943}\inst{\ref{aff9}}
\and H.~J.~Dickinson\orcid{0000-0003-0475-008X}\inst{\ref{aff9}}
\and L.~Fortson\orcid{0000-0002-1067-8558}\inst{\ref{aff10}}
\and I.~Garland\orcid{0000-0002-3887-6433}\inst{\ref{aff11}}
\and T.~G\'eron\orcid{0000-0002-6851-9613}\inst{\ref{aff5}}
\and W.~Keel\inst{\ref{aff12}}
\and S.~Kruk\orcid{0000-0001-8010-8879}\inst{\ref{aff13}}
\and C.~J.~Lintott\orcid{0000-0001-5578-359X}\inst{\ref{aff14}}
\and K.~Mantha\orcid{0000-0002-6016-300X}\inst{\ref{aff10}}
\and K.~Masters\orcid{0000-0003-0846-9578}\inst{\ref{aff15}}
\and D.~O'Ryan\orcid{0000-0003-1217-4617}\inst{\ref{aff16}}
\and J.~J.~Popp\orcid{0000-0002-3724-1727}\inst{\ref{aff9}}
\and H.~Roberts\orcid{0000-0003-0046-9848}\inst{\ref{aff10}}
\and C.~Scarlata\orcid{0000-0002-9136-8876}\inst{\ref{aff10}}
\and J.~S.~Makechemu\orcid{0009-0009-6545-8710}\inst{\ref{aff17}}
\and B.~Simmons\orcid{0000-0001-5882-3323}\inst{\ref{aff17}}
\and R.~J.~Smethurst\orcid{0000-0001-6417-7196}\inst{\ref{aff14}}
\and A.~Spindler\orcid{0000-0003-0198-3881}\inst{\ref{aff18}}
\and M.~Baes\orcid{0000-0002-3930-2757}\inst{\ref{aff19}}
\and E.~M.~Corsini\orcid{0000-0003-3460-5633}\inst{\ref{aff20},\ref{aff21}}
\and H.~Dom\'inguez~S\'anchez\orcid{0000-0002-9013-1316}\inst{\ref{aff22}}
\and E.~Duran-Camacho\orcid{0000-0002-3153-0536}\inst{\ref{aff1}}
\and H.~Fu\orcid{0009-0002-8051-1056}\inst{\ref{aff23},\ref{aff24}}
\and J.~Junais\orcid{0000-0002-7016-4532}\inst{\ref{aff1},\ref{aff8}}
\and J.~Mendez-Abreu\orcid{0000-0002-8766-2597}\inst{\ref{aff8},\ref{aff1}}
\and A.~Nersesian\orcid{0000-0001-6843-409X}\inst{\ref{aff25},\ref{aff19}}
\and F.~Shankar\orcid{0000-0001-8973-5051}\inst{\ref{aff24}}
\and M.~N.~Le\orcid{0009-0003-0674-9813}\inst{\ref{aff1},\ref{aff8}}
\and J.~Vega-Ferrero\orcid{0000-0003-2338-5567}\inst{\ref{aff26}}
\and L.~Wang\orcid{0000-0002-6736-9158}\inst{\ref{aff27},\ref{aff28}}
\and N.~Aghanim\orcid{0000-0002-6688-8992}\inst{\ref{aff29}}
\and B.~Altieri\orcid{0000-0003-3936-0284}\inst{\ref{aff13}}
\and A.~Amara\inst{\ref{aff30}}
\and S.~Andreon\orcid{0000-0002-2041-8784}\inst{\ref{aff31}}
\and N.~Auricchio\orcid{0000-0003-4444-8651}\inst{\ref{aff32}}
\and C.~Baccigalupi\orcid{0000-0002-8211-1630}\inst{\ref{aff33},\ref{aff34},\ref{aff35},\ref{aff36}}
\and M.~Baldi\orcid{0000-0003-4145-1943}\inst{\ref{aff37},\ref{aff32},\ref{aff38}}
\and A.~Balestra\orcid{0000-0002-6967-261X}\inst{\ref{aff21}}
\and S.~Bardelli\orcid{0000-0002-8900-0298}\inst{\ref{aff32}}
\and A.~Basset\inst{\ref{aff39}}
\and P.~Battaglia\orcid{0000-0002-7337-5909}\inst{\ref{aff32}}
\and F.~Bernardeau\inst{\ref{aff40},\ref{aff41}}
\and A.~Biviano\orcid{0000-0002-0857-0732}\inst{\ref{aff34},\ref{aff33}}
\and A.~Bonchi\orcid{0000-0002-2667-5482}\inst{\ref{aff42}}
\and E.~Branchini\orcid{0000-0002-0808-6908}\inst{\ref{aff43},\ref{aff44},\ref{aff31}}
\and M.~Brescia\orcid{0000-0001-9506-5680}\inst{\ref{aff45},\ref{aff46}}
\and J.~Brinchmann\orcid{0000-0003-4359-8797}\inst{\ref{aff47},\ref{aff48}}
\and S.~Camera\orcid{0000-0003-3399-3574}\inst{\ref{aff49},\ref{aff50},\ref{aff51}}
\and V.~Capobianco\orcid{0000-0002-3309-7692}\inst{\ref{aff51}}
\and C.~Carbone\orcid{0000-0003-0125-3563}\inst{\ref{aff52}}
\and J.~Carretero\orcid{0000-0002-3130-0204}\inst{\ref{aff53},\ref{aff54}}
\and S.~Casas\orcid{0000-0002-4751-5138}\inst{\ref{aff55}}
\and M.~Castellano\orcid{0000-0001-9875-8263}\inst{\ref{aff56}}
\and G.~Castignani\orcid{0000-0001-6831-0687}\inst{\ref{aff32}}
\and S.~Cavuoti\orcid{0000-0002-3787-4196}\inst{\ref{aff46},\ref{aff57}}
\and K.~C.~Chambers\orcid{0000-0001-6965-7789}\inst{\ref{aff58}}
\and A.~Cimatti\inst{\ref{aff59}}
\and C.~Colodro-Conde\inst{\ref{aff1}}
\and G.~Congedo\orcid{0000-0003-2508-0046}\inst{\ref{aff60}}
\and C.~J.~Conselice\orcid{0000-0003-1949-7638}\inst{\ref{aff6}}
\and L.~Conversi\orcid{0000-0002-6710-8476}\inst{\ref{aff61},\ref{aff13}}
\and Y.~Copin\orcid{0000-0002-5317-7518}\inst{\ref{aff62}}
\and F.~Courbin\orcid{0000-0003-0758-6510}\inst{\ref{aff63},\ref{aff64}}
\and H.~M.~Courtois\orcid{0000-0003-0509-1776}\inst{\ref{aff65}}
\and M.~Cropper\orcid{0000-0003-4571-9468}\inst{\ref{aff66}}
\and A.~Da~Silva\orcid{0000-0002-6385-1609}\inst{\ref{aff67},\ref{aff68}}
\and H.~Degaudenzi\orcid{0000-0002-5887-6799}\inst{\ref{aff69}}
\and G.~De~Lucia\orcid{0000-0002-6220-9104}\inst{\ref{aff34}}
\and A.~M.~Di~Giorgio\orcid{0000-0002-4767-2360}\inst{\ref{aff70}}
\and C.~Dolding\orcid{0009-0003-7199-6108}\inst{\ref{aff66}}
\and H.~Dole\orcid{0000-0002-9767-3839}\inst{\ref{aff29}}
\and F.~Dubath\orcid{0000-0002-6533-2810}\inst{\ref{aff69}}
\and C.~A.~J.~Duncan\orcid{0009-0003-3573-0791}\inst{\ref{aff6}}
\and X.~Dupac\inst{\ref{aff13}}
\and S.~Dusini\orcid{0000-0002-1128-0664}\inst{\ref{aff71}}
\and A.~Ealet\orcid{0000-0003-3070-014X}\inst{\ref{aff62}}
\and S.~Escoffier\orcid{0000-0002-2847-7498}\inst{\ref{aff72}}
\and M.~Fabricius\orcid{0000-0002-7025-6058}\inst{\ref{aff73},\ref{aff74}}
\and M.~Farina\orcid{0000-0002-3089-7846}\inst{\ref{aff70}}
\and R.~Farinelli\inst{\ref{aff32}}
\and F.~Faustini\orcid{0000-0001-6274-5145}\inst{\ref{aff42},\ref{aff56}}
\and S.~Ferriol\inst{\ref{aff62}}
\and F.~Finelli\orcid{0000-0002-6694-3269}\inst{\ref{aff32},\ref{aff75}}
\and S.~Fotopoulou\orcid{0000-0002-9686-254X}\inst{\ref{aff76}}
\and M.~Frailis\orcid{0000-0002-7400-2135}\inst{\ref{aff34}}
\and S.~Galeotta\orcid{0000-0002-3748-5115}\inst{\ref{aff34}}
\and K.~George\orcid{0000-0002-1734-8455}\inst{\ref{aff74}}
\and W.~Gillard\orcid{0000-0003-4744-9748}\inst{\ref{aff72}}
\and B.~Gillis\orcid{0000-0002-4478-1270}\inst{\ref{aff60}}
\and C.~Giocoli\orcid{0000-0002-9590-7961}\inst{\ref{aff32},\ref{aff38}}
\and J.~Gracia-Carpio\inst{\ref{aff73}}
\and A.~Grazian\orcid{0000-0002-5688-0663}\inst{\ref{aff21}}
\and F.~Grupp\inst{\ref{aff73},\ref{aff74}}
\and S.~Gwyn\orcid{0000-0001-8221-8406}\inst{\ref{aff77}}
\and S.~V.~H.~Haugan\orcid{0000-0001-9648-7260}\inst{\ref{aff78}}
\and H.~Hoekstra\orcid{0000-0002-0641-3231}\inst{\ref{aff79}}
\and W.~Holmes\inst{\ref{aff80}}
\and I.~M.~Hook\orcid{0000-0002-2960-978X}\inst{\ref{aff17}}
\and F.~Hormuth\inst{\ref{aff81}}
\and A.~Hornstrup\orcid{0000-0002-3363-0936}\inst{\ref{aff82},\ref{aff83}}
\and P.~Hudelot\inst{\ref{aff41}}
\and K.~Jahnke\orcid{0000-0003-3804-2137}\inst{\ref{aff84}}
\and M.~Jhabvala\inst{\ref{aff85}}
\and E.~Keih\"anen\orcid{0000-0003-1804-7715}\inst{\ref{aff86}}
\and S.~Kermiche\orcid{0000-0002-0302-5735}\inst{\ref{aff72}}
\and B.~Kubik\orcid{0009-0006-5823-4880}\inst{\ref{aff62}}
\and K.~Kuijken\orcid{0000-0002-3827-0175}\inst{\ref{aff79}}
\and M.~K\"ummel\orcid{0000-0003-2791-2117}\inst{\ref{aff74}}
\and M.~Kunz\orcid{0000-0002-3052-7394}\inst{\ref{aff87}}
\and H.~Kurki-Suonio\orcid{0000-0002-4618-3063}\inst{\ref{aff88},\ref{aff89}}
\and Q.~Le~Boulc'h\inst{\ref{aff90}}
\and A.~M.~C.~Le~Brun\orcid{0000-0002-0936-4594}\inst{\ref{aff91}}
\and D.~Le~Mignant\orcid{0000-0002-5339-5515}\inst{\ref{aff92}}
\and S.~Ligori\orcid{0000-0003-4172-4606}\inst{\ref{aff51}}
\and P.~B.~Lilje\orcid{0000-0003-4324-7794}\inst{\ref{aff78}}
\and V.~Lindholm\orcid{0000-0003-2317-5471}\inst{\ref{aff88},\ref{aff89}}
\and I.~Lloro\orcid{0000-0001-5966-1434}\inst{\ref{aff93}}
\and D.~Maino\inst{\ref{aff94},\ref{aff52},\ref{aff95}}
\and E.~Maiorano\orcid{0000-0003-2593-4355}\inst{\ref{aff32}}
\and O.~Mansutti\orcid{0000-0001-5758-4658}\inst{\ref{aff34}}
\and S.~Marcin\inst{\ref{aff96}}
\and O.~Marggraf\orcid{0000-0001-7242-3852}\inst{\ref{aff97}}
\and M.~Martinelli\orcid{0000-0002-6943-7732}\inst{\ref{aff56},\ref{aff98}}
\and N.~Martinet\orcid{0000-0003-2786-7790}\inst{\ref{aff92}}
\and F.~Marulli\orcid{0000-0002-8850-0303}\inst{\ref{aff99},\ref{aff32},\ref{aff38}}
\and R.~Massey\orcid{0000-0002-6085-3780}\inst{\ref{aff100}}
\and H.~J.~McCracken\orcid{0000-0002-9489-7765}\inst{\ref{aff41}}
\and E.~Medinaceli\orcid{0000-0002-4040-7783}\inst{\ref{aff32}}
\and M.~Melchior\inst{\ref{aff101}}
\and Y.~Mellier\inst{\ref{aff102},\ref{aff41}}
\and M.~Meneghetti\orcid{0000-0003-1225-7084}\inst{\ref{aff32},\ref{aff38}}
\and E.~Merlin\orcid{0000-0001-6870-8900}\inst{\ref{aff56}}
\and G.~Meylan\inst{\ref{aff103}}
\and A.~Mora\orcid{0000-0002-1922-8529}\inst{\ref{aff104}}
\and M.~Moresco\orcid{0000-0002-7616-7136}\inst{\ref{aff99},\ref{aff32}}
\and L.~Moscardini\orcid{0000-0002-3473-6716}\inst{\ref{aff99},\ref{aff32},\ref{aff38}}
\and C.~Neissner\orcid{0000-0001-8524-4968}\inst{\ref{aff105},\ref{aff54}}
\and R.~C.~Nichol\orcid{0000-0003-0939-6518}\inst{\ref{aff30}}
\and S.-M.~Niemi\inst{\ref{aff106}}
\and J.~W.~Nightingale\orcid{0000-0002-8987-7401}\inst{\ref{aff107}}
\and C.~Padilla\orcid{0000-0001-7951-0166}\inst{\ref{aff105}}
\and S.~Paltani\orcid{0000-0002-8108-9179}\inst{\ref{aff69}}
\and F.~Pasian\orcid{0000-0002-4869-3227}\inst{\ref{aff34}}
\and K.~Pedersen\inst{\ref{aff108}}
\and W.~J.~Percival\orcid{0000-0002-0644-5727}\inst{\ref{aff109},\ref{aff110},\ref{aff111}}
\and V.~Pettorino\inst{\ref{aff106}}
\and S.~Pires\orcid{0000-0002-0249-2104}\inst{\ref{aff112}}
\and G.~Polenta\orcid{0000-0003-4067-9196}\inst{\ref{aff42}}
\and M.~Poncet\inst{\ref{aff39}}
\and L.~A.~Popa\inst{\ref{aff113}}
\and L.~Pozzetti\orcid{0000-0001-7085-0412}\inst{\ref{aff32}}
\and F.~Raison\orcid{0000-0002-7819-6918}\inst{\ref{aff73}}
\and A.~Renzi\orcid{0000-0001-9856-1970}\inst{\ref{aff114},\ref{aff71}}
\and J.~Rhodes\orcid{0000-0002-4485-8549}\inst{\ref{aff80}}
\and G.~Riccio\inst{\ref{aff46}}
\and E.~Romelli\orcid{0000-0003-3069-9222}\inst{\ref{aff34}}
\and M.~Roncarelli\orcid{0000-0001-9587-7822}\inst{\ref{aff32}}
\and R.~Saglia\orcid{0000-0003-0378-7032}\inst{\ref{aff74},\ref{aff73}}
\and Z.~Sakr\orcid{0000-0002-4823-3757}\inst{\ref{aff115},\ref{aff116},\ref{aff117}}
\and D.~Sapone\orcid{0000-0001-7089-4503}\inst{\ref{aff118}}
\and B.~Sartoris\orcid{0000-0003-1337-5269}\inst{\ref{aff74},\ref{aff34}}
\and M.~Schirmer\orcid{0000-0003-2568-9994}\inst{\ref{aff84}}
\and P.~Schneider\orcid{0000-0001-8561-2679}\inst{\ref{aff97}}
\and M.~Scodeggio\inst{\ref{aff52}}
\and A.~Secroun\orcid{0000-0003-0505-3710}\inst{\ref{aff72}}
\and G.~Seidel\orcid{0000-0003-2907-353X}\inst{\ref{aff84}}
\and M.~Seiffert\orcid{0000-0002-7536-9393}\inst{\ref{aff80}}
\and S.~Serrano\orcid{0000-0002-0211-2861}\inst{\ref{aff119},\ref{aff120},\ref{aff7}}
\and P.~Simon\inst{\ref{aff97}}
\and C.~Sirignano\orcid{0000-0002-0995-7146}\inst{\ref{aff114},\ref{aff71}}
\and G.~Sirri\orcid{0000-0003-2626-2853}\inst{\ref{aff38}}
\and L.~Stanco\orcid{0000-0002-9706-5104}\inst{\ref{aff71}}
\and J.~Steinwagner\orcid{0000-0001-7443-1047}\inst{\ref{aff73}}
\and P.~Tallada-Cresp\'{i}\orcid{0000-0002-1336-8328}\inst{\ref{aff53},\ref{aff54}}
\and A.~N.~Taylor\inst{\ref{aff60}}
\and I.~Tereno\inst{\ref{aff67},\ref{aff121}}
\and S.~Toft\orcid{0000-0003-3631-7176}\inst{\ref{aff122},\ref{aff123}}
\and R.~Toledo-Moreo\orcid{0000-0002-2997-4859}\inst{\ref{aff124}}
\and F.~Torradeflot\orcid{0000-0003-1160-1517}\inst{\ref{aff54},\ref{aff53}}
\and I.~Tutusaus\orcid{0000-0002-3199-0399}\inst{\ref{aff116}}
\and L.~Valenziano\orcid{0000-0002-1170-0104}\inst{\ref{aff32},\ref{aff75}}
\and J.~Valiviita\orcid{0000-0001-6225-3693}\inst{\ref{aff88},\ref{aff89}}
\and T.~Vassallo\orcid{0000-0001-6512-6358}\inst{\ref{aff74},\ref{aff34}}
\and G.~Verdoes~Kleijn\orcid{0000-0001-5803-2580}\inst{\ref{aff28}}
\and Y.~Wang\orcid{0000-0002-4749-2984}\inst{\ref{aff125}}
\and J.~Weller\orcid{0000-0002-8282-2010}\inst{\ref{aff74},\ref{aff73}}
\and A.~Zacchei\orcid{0000-0003-0396-1192}\inst{\ref{aff34},\ref{aff33}}
\and G.~Zamorani\orcid{0000-0002-2318-301X}\inst{\ref{aff32}}
\and F.~M.~Zerbi\inst{\ref{aff31}}
\and I.~A.~Zinchenko\orcid{0000-0002-2944-2449}\inst{\ref{aff74}}
\and E.~Zucca\orcid{0000-0002-5845-8132}\inst{\ref{aff32}}
\and V.~Allevato\orcid{0000-0001-7232-5152}\inst{\ref{aff46}}
\and M.~Ballardini\orcid{0000-0003-4481-3559}\inst{\ref{aff126},\ref{aff127},\ref{aff32}}
\and M.~Bolzonella\orcid{0000-0003-3278-4607}\inst{\ref{aff32}}
\and E.~Bozzo\orcid{0000-0002-8201-1525}\inst{\ref{aff69}}
\and C.~Burigana\orcid{0000-0002-3005-5796}\inst{\ref{aff128},\ref{aff75}}
\and A.~Cappi\inst{\ref{aff32},\ref{aff129}}
\and D.~Di~Ferdinando\inst{\ref{aff38}}
\and J.~A.~Escartin~Vigo\inst{\ref{aff73}}
\and L.~Gabarra\orcid{0000-0002-8486-8856}\inst{\ref{aff14}}
\and J.~Mart\'{i}n-Fleitas\orcid{0000-0002-8594-569X}\inst{\ref{aff104}}
\and S.~Matthew\orcid{0000-0001-8448-1697}\inst{\ref{aff60}}
\and N.~Mauri\orcid{0000-0001-8196-1548}\inst{\ref{aff59},\ref{aff38}}
\and R.~B.~Metcalf\orcid{0000-0003-3167-2574}\inst{\ref{aff99},\ref{aff32}}
\and A.~Pezzotta\orcid{0000-0003-0726-2268}\inst{\ref{aff73}}
\and M.~P\"ontinen\orcid{0000-0001-5442-2530}\inst{\ref{aff88}}
\and C.~Porciani\orcid{0000-0002-7797-2508}\inst{\ref{aff97}}
\and I.~Risso\orcid{0000-0003-2525-7761}\inst{\ref{aff130}}
\and V.~Scottez\inst{\ref{aff102},\ref{aff131}}
\and M.~Sereno\orcid{0000-0003-0302-0325}\inst{\ref{aff32},\ref{aff38}}
\and M.~Tenti\orcid{0000-0002-4254-5901}\inst{\ref{aff38}}
\and M.~Viel\orcid{0000-0002-2642-5707}\inst{\ref{aff33},\ref{aff34},\ref{aff36},\ref{aff35},\ref{aff132}}
\and M.~Wiesmann\orcid{0009-0000-8199-5860}\inst{\ref{aff78}}
\and Y.~Akrami\orcid{0000-0002-2407-7956}\inst{\ref{aff133},\ref{aff134}}
\and I.~T.~Andika\orcid{0000-0001-6102-9526}\inst{\ref{aff135},\ref{aff136}}
\and S.~Anselmi\orcid{0000-0002-3579-9583}\inst{\ref{aff71},\ref{aff114},\ref{aff137}}
\and M.~Archidiacono\orcid{0000-0003-4952-9012}\inst{\ref{aff94},\ref{aff95}}
\and F.~Atrio-Barandela\orcid{0000-0002-2130-2513}\inst{\ref{aff138}}
\and C.~Benoist\inst{\ref{aff129}}
\and K.~Benson\inst{\ref{aff66}}
\and D.~Bertacca\orcid{0000-0002-2490-7139}\inst{\ref{aff114},\ref{aff21},\ref{aff71}}
\and M.~Bethermin\orcid{0000-0002-3915-2015}\inst{\ref{aff139}}
\and L.~Bisigello\orcid{0000-0003-0492-4924}\inst{\ref{aff21}}
\and A.~Blanchard\orcid{0000-0001-8555-9003}\inst{\ref{aff116}}
\and L.~Blot\orcid{0000-0002-9622-7167}\inst{\ref{aff140},\ref{aff137}}
\and H.~B\"ohringer\orcid{0000-0001-8241-4204}\inst{\ref{aff73},\ref{aff141},\ref{aff142}}
\and S.~Borgani\orcid{0000-0001-6151-6439}\inst{\ref{aff143},\ref{aff33},\ref{aff34},\ref{aff35},\ref{aff132}}
\and M.~L.~Brown\orcid{0000-0002-0370-8077}\inst{\ref{aff6}}
\and S.~Bruton\orcid{0000-0002-6503-5218}\inst{\ref{aff144}}
\and A.~Calabro\orcid{0000-0003-2536-1614}\inst{\ref{aff56}}
\and B.~Camacho~Quevedo\orcid{0000-0002-8789-4232}\inst{\ref{aff119},\ref{aff7}}
\and F.~Caro\inst{\ref{aff56}}
\and C.~S.~Carvalho\inst{\ref{aff121}}
\and T.~Castro\orcid{0000-0002-6292-3228}\inst{\ref{aff34},\ref{aff35},\ref{aff33},\ref{aff132}}
\and Y.~Charles\inst{\ref{aff92}}
\and F.~Cogato\orcid{0000-0003-4632-6113}\inst{\ref{aff99},\ref{aff32}}
\and T.~Contini\orcid{0000-0003-0275-938X}\inst{\ref{aff116}}
\and A.~R.~Cooray\orcid{0000-0002-3892-0190}\inst{\ref{aff145}}
\and O.~Cucciati\orcid{0000-0002-9336-7551}\inst{\ref{aff32}}
\and S.~Davini\orcid{0000-0003-3269-1718}\inst{\ref{aff44}}
\and F.~De~Paolis\orcid{0000-0001-6460-7563}\inst{\ref{aff146},\ref{aff147},\ref{aff148}}
\and G.~Desprez\orcid{0000-0001-8325-1742}\inst{\ref{aff28}}
\and A.~D\'iaz-S\'anchez\orcid{0000-0003-0748-4768}\inst{\ref{aff149}}
\and J.~J.~Diaz\inst{\ref{aff1}}
\and S.~Di~Domizio\orcid{0000-0003-2863-5895}\inst{\ref{aff43},\ref{aff44}}
\and J.~M.~Diego\orcid{0000-0001-9065-3926}\inst{\ref{aff22}}
\and P.-A.~Duc\orcid{0000-0003-3343-6284}\inst{\ref{aff139}}
\and A.~Enia\orcid{0000-0002-0200-2857}\inst{\ref{aff37},\ref{aff32}}
\and Y.~Fang\inst{\ref{aff74}}
\and A.~G.~Ferrari\orcid{0009-0005-5266-4110}\inst{\ref{aff38}}
\and A.~Finoguenov\orcid{0000-0002-4606-5403}\inst{\ref{aff88}}
\and A.~Fontana\orcid{0000-0003-3820-2823}\inst{\ref{aff56}}
\and F.~Fontanot\orcid{0000-0003-4744-0188}\inst{\ref{aff34},\ref{aff33}}
\and A.~Franco\orcid{0000-0002-4761-366X}\inst{\ref{aff147},\ref{aff146},\ref{aff148}}
\and K.~Ganga\orcid{0000-0001-8159-8208}\inst{\ref{aff150}}
\and J.~Garc\'ia-Bellido\orcid{0000-0002-9370-8360}\inst{\ref{aff133}}
\and T.~Gasparetto\orcid{0000-0002-7913-4866}\inst{\ref{aff34}}
\and V.~Gautard\inst{\ref{aff151}}
\and E.~Gaztanaga\orcid{0000-0001-9632-0815}\inst{\ref{aff7},\ref{aff119},\ref{aff152}}
\and F.~Giacomini\orcid{0000-0002-3129-2814}\inst{\ref{aff38}}
\and G.~Gozaliasl\orcid{0000-0002-0236-919X}\inst{\ref{aff153},\ref{aff88}}
\and M.~Guidi\orcid{0000-0001-9408-1101}\inst{\ref{aff37},\ref{aff32}}
\and C.~M.~Gutierrez\orcid{0000-0001-7854-783X}\inst{\ref{aff154}}
\and A.~Hall\orcid{0000-0002-3139-8651}\inst{\ref{aff60}}
\and W.~G.~Hartley\inst{\ref{aff69}}
\and S.~Hemmati\orcid{0000-0003-2226-5395}\inst{\ref{aff155}}
\and C.~Hern\'andez-Monteagudo\orcid{0000-0001-5471-9166}\inst{\ref{aff8},\ref{aff1}}
\and H.~Hildebrandt\orcid{0000-0002-9814-3338}\inst{\ref{aff156}}
\and J.~Hjorth\orcid{0000-0002-4571-2306}\inst{\ref{aff108}}
\and J.~J.~E.~Kajava\orcid{0000-0002-3010-8333}\inst{\ref{aff157},\ref{aff158}}
\and Y.~Kang\orcid{0009-0000-8588-7250}\inst{\ref{aff69}}
\and V.~Kansal\orcid{0000-0002-4008-6078}\inst{\ref{aff159},\ref{aff160}}
\and D.~Karagiannis\orcid{0000-0002-4927-0816}\inst{\ref{aff126},\ref{aff161}}
\and K.~Kiiveri\inst{\ref{aff86}}
\and C.~C.~Kirkpatrick\inst{\ref{aff86}}
\and J.~Le~Graet\orcid{0000-0001-6523-7971}\inst{\ref{aff72}}
\and L.~Legrand\orcid{0000-0003-0610-5252}\inst{\ref{aff162},\ref{aff163}}
\and M.~Lembo\orcid{0000-0002-5271-5070}\inst{\ref{aff126},\ref{aff127}}
\and F.~Lepori\orcid{0009-0000-5061-7138}\inst{\ref{aff164}}
\and G.~Leroy\orcid{0009-0004-2523-4425}\inst{\ref{aff165},\ref{aff100}}
\and G.~F.~Lesci\orcid{0000-0002-4607-2830}\inst{\ref{aff99},\ref{aff32}}
\and J.~Lesgourgues\orcid{0000-0001-7627-353X}\inst{\ref{aff55}}
\and L.~Leuzzi\orcid{0009-0006-4479-7017}\inst{\ref{aff99},\ref{aff32}}
\and T.~I.~Liaudat\orcid{0000-0002-9104-314X}\inst{\ref{aff166}}
\and A.~Loureiro\orcid{0000-0002-4371-0876}\inst{\ref{aff167},\ref{aff168}}
\and J.~Macias-Perez\orcid{0000-0002-5385-2763}\inst{\ref{aff169}}
\and G.~Maggio\orcid{0000-0003-4020-4836}\inst{\ref{aff34}}
\and M.~Magliocchetti\orcid{0000-0001-9158-4838}\inst{\ref{aff70}}
\and E.~A.~Magnier\orcid{0000-0002-7965-2815}\inst{\ref{aff58}}
\and F.~Mannucci\orcid{0000-0002-4803-2381}\inst{\ref{aff170}}
\and R.~Maoli\orcid{0000-0002-6065-3025}\inst{\ref{aff171},\ref{aff56}}
\and C.~J.~A.~P.~Martins\orcid{0000-0002-4886-9261}\inst{\ref{aff172},\ref{aff47}}
\and L.~Maurin\orcid{0000-0002-8406-0857}\inst{\ref{aff29}}
\and M.~Miluzio\inst{\ref{aff13},\ref{aff173}}
\and P.~Monaco\orcid{0000-0003-2083-7564}\inst{\ref{aff143},\ref{aff34},\ref{aff35},\ref{aff33}}
\and C.~Moretti\orcid{0000-0003-3314-8936}\inst{\ref{aff36},\ref{aff132},\ref{aff34},\ref{aff33},\ref{aff35}}
\and G.~Morgante\inst{\ref{aff32}}
\and C.~Murray\inst{\ref{aff150}}
\and K.~Naidoo\orcid{0000-0002-9182-1802}\inst{\ref{aff152}}
\and A.~Navarro-Alsina\orcid{0000-0002-3173-2592}\inst{\ref{aff97}}
\and S.~Nesseris\orcid{0000-0002-0567-0324}\inst{\ref{aff133}}
\and F.~Passalacqua\orcid{0000-0002-8606-4093}\inst{\ref{aff114},\ref{aff71}}
\and K.~Paterson\orcid{0000-0001-8340-3486}\inst{\ref{aff84}}
\and L.~Patrizii\inst{\ref{aff38}}
\and A.~Pisani\orcid{0000-0002-6146-4437}\inst{\ref{aff72},\ref{aff174}}
\and D.~Potter\orcid{0000-0002-0757-5195}\inst{\ref{aff164}}
\and S.~Quai\orcid{0000-0002-0449-8163}\inst{\ref{aff99},\ref{aff32}}
\and M.~Radovich\orcid{0000-0002-3585-866X}\inst{\ref{aff21}}
\and G.~Rodighiero\orcid{0000-0002-9415-2296}\inst{\ref{aff114},\ref{aff21}}
\and S.~Sacquegna\orcid{0000-0002-8433-6630}\inst{\ref{aff146},\ref{aff147},\ref{aff148}}
\and M.~Sahl\'en\orcid{0000-0003-0973-4804}\inst{\ref{aff175}}
\and D.~B.~Sanders\orcid{0000-0002-1233-9998}\inst{\ref{aff58}}
\and E.~Sarpa\orcid{0000-0002-1256-655X}\inst{\ref{aff36},\ref{aff132},\ref{aff35}}
\and J.~Schaye\orcid{0000-0002-0668-5560}\inst{\ref{aff79}}
\and A.~Schneider\orcid{0000-0001-7055-8104}\inst{\ref{aff164}}
\and M.~Schultheis\inst{\ref{aff129}}
\and D.~Sciotti\orcid{0009-0008-4519-2620}\inst{\ref{aff56},\ref{aff98}}
\and E.~Sellentin\inst{\ref{aff176},\ref{aff79}}
\and L.~C.~Smith\orcid{0000-0002-3259-2771}\inst{\ref{aff177}}
\and K.~Tanidis\orcid{0000-0001-9843-5130}\inst{\ref{aff14}}
\and G.~Testera\inst{\ref{aff44}}
\and R.~Teyssier\orcid{0000-0001-7689-0933}\inst{\ref{aff174}}
\and S.~Tosi\orcid{0000-0002-7275-9193}\inst{\ref{aff43},\ref{aff130}}
\and A.~Troja\orcid{0000-0003-0239-4595}\inst{\ref{aff114},\ref{aff71}}
\and M.~Tucci\inst{\ref{aff69}}
\and C.~Valieri\inst{\ref{aff38}}
\and A.~Venhola\orcid{0000-0001-6071-4564}\inst{\ref{aff178}}
\and D.~Vergani\orcid{0000-0003-0898-2216}\inst{\ref{aff32}}
\and G.~Verza\orcid{0000-0002-1886-8348}\inst{\ref{aff179}}
\and P.~Vielzeuf\orcid{0000-0003-2035-9339}\inst{\ref{aff72}}
\and N.~A.~Walton\orcid{0000-0003-3983-8778}\inst{\ref{aff177}}
\and E.~Soubrie\orcid{0000-0001-9295-1863}\inst{\ref{aff29}}
\and D.~Scott\orcid{0000-0002-6878-9840}\inst{\ref{aff180}}}
										   
\institute{Instituto de Astrof\'{\i}sica de Canarias, V\'{\i}a L\'actea, 38205 La Laguna, Tenerife, Spain\label{aff1}
\and
Instituto de Astrof\'isica de Canarias (IAC); Departamento de Astrof\'isica, Universidad de La Laguna (ULL), 38200, La Laguna, Tenerife, Spain\label{aff2}
\and
Universit\'e PSL, Observatoire de Paris, Sorbonne Universit\'e, CNRS, LERMA, 75014, Paris, France\label{aff3}
\and
Universit\'e Paris-Cit\'e, 5 Rue Thomas Mann, 75013, Paris, France\label{aff4}
\and
David A. Dunlap Department of Astronomy \& Astrophysics, University of Toronto, 50 St George Street, Toronto, Ontario M5S 3H4, Canada\label{aff5}
\and
Jodrell Bank Centre for Astrophysics, Department of Physics and Astronomy, University of Manchester, Oxford Road, Manchester M13 9PL, UK\label{aff6}
\and
Institute of Space Sciences (ICE, CSIC), Campus UAB, Carrer de Can Magrans, s/n, 08193 Barcelona, Spain\label{aff7}
\and
Universidad de La Laguna, Departamento de Astrof\'{\i}sica, 38206 La Laguna, Tenerife, Spain\label{aff8}
\and
School of Physical Sciences, The Open University, Milton Keynes, MK7 6AA, UK\label{aff9}
\and
Minnesota Institute for Astrophysics, University of Minnesota, 116 Church St SE, Minneapolis, MN 55455, USA\label{aff10}
\and
Masaryk University, Kotl\'{a}\v{r}sk\'{a} 2, Brno, 611 37, Czech Republic\label{aff11}
\and
University of Alabama, Tuscaloosa, AL 35487, USA\label{aff12}
\and
ESAC/ESA, Camino Bajo del Castillo, s/n., Urb. Villafranca del Castillo, 28692 Villanueva de la Ca\~nada, Madrid, Spain\label{aff13}
\and
Department of Physics, Oxford University, Keble Road, Oxford OX1 3RH, UK\label{aff14}
\and
Departments of Physics and Astronomy, Haverford College, 370 Lancaster Avenue, Haverford, PA 19041, USA\label{aff15}
\and
Centro de Astrobiolog\'ia (CAB), CSIC-INTA, ESAC Campus, Camino Bajo del Castillo s/n, 28692 Villanueva de la Ca\~nada, Madrid, Spain\label{aff16}
\and
Department of Physics, Lancaster University, Lancaster, LA1 4YB, UK\label{aff17}
\and
School of Physics, Astronomy and Mathematics, University of Hertfordshire, College Lane, Hatfield AL10 9AB, UK\label{aff18}
\and
Sterrenkundig Observatorium, Universiteit Gent, Krijgslaan 281 S9, 9000 Gent, Belgium\label{aff19}
\and
Dipartimento di Fisica e Astronomia ``G. Galilei", Universit\`a di Padova, Vicolo dell'Osservatorio 3, 35122 Padova, Italy\label{aff20}
\and
INAF-Osservatorio Astronomico di Padova, Via dell'Osservatorio 5, 35122 Padova, Italy\label{aff21}
\and
Instituto de F\'isica de Cantabria, Edificio Juan Jord\'a, Avenida de los Castros, 39005 Santander, Spain\label{aff22}
\and
Center for Astronomy and Astrophysics and Department of Physics, Fudan University, Shanghai 200438, People's Republic of China\label{aff23}
\and
School of Physics \& Astronomy, University of Southampton, Highfield Campus, Southampton SO17 1BJ, UK\label{aff24}
\and
STAR Institute, University of Li{\`e}ge, Quartier Agora, All\'ee du six Ao\^ut 19c, 4000 Li\`ege, Belgium\label{aff25}
\and
Centro de Estudios de F\'isica del Cosmos de Arag\'on (CEFCA), Plaza San Juan, 1, planta 2, 44001, Teruel, Spain\label{aff26}
\and
SRON Netherlands Institute for Space Research, Landleven 12, 9747 AD, Groningen, The Netherlands\label{aff27}
\and
Kapteyn Astronomical Institute, University of Groningen, PO Box 800, 9700 AV Groningen, The Netherlands\label{aff28}
\and
Universit\'e Paris-Saclay, CNRS, Institut d'astrophysique spatiale, 91405, Orsay, France\label{aff29}
\and
School of Mathematics and Physics, University of Surrey, Guildford, Surrey, GU2 7XH, UK\label{aff30}
\and
INAF-Osservatorio Astronomico di Brera, Via Brera 28, 20122 Milano, Italy\label{aff31}
\and
INAF-Osservatorio di Astrofisica e Scienza dello Spazio di Bologna, Via Piero Gobetti 93/3, 40129 Bologna, Italy\label{aff32}
\and
IFPU, Institute for Fundamental Physics of the Universe, via Beirut 2, 34151 Trieste, Italy\label{aff33}
\and
INAF-Osservatorio Astronomico di Trieste, Via G. B. Tiepolo 11, 34143 Trieste, Italy\label{aff34}
\and
INFN, Sezione di Trieste, Via Valerio 2, 34127 Trieste TS, Italy\label{aff35}
\and
SISSA, International School for Advanced Studies, Via Bonomea 265, 34136 Trieste TS, Italy\label{aff36}
\and
Dipartimento di Fisica e Astronomia, Universit\`a di Bologna, Via Gobetti 93/2, 40129 Bologna, Italy\label{aff37}
\and
INFN-Sezione di Bologna, Viale Berti Pichat 6/2, 40127 Bologna, Italy\label{aff38}
\and
Centre National d'Etudes Spatiales -- Centre spatial de Toulouse, 18 avenue Edouard Belin, 31401 Toulouse Cedex 9, France\label{aff39}
\and
Institut de Physique Th\'eorique, CEA, CNRS, Universit\'e Paris-Saclay 91191 Gif-sur-Yvette Cedex, France\label{aff40}
\and
Institut d'Astrophysique de Paris, UMR 7095, CNRS, and Sorbonne Universit\'e, 98 bis boulevard Arago, 75014 Paris, France\label{aff41}
\and
Space Science Data Center, Italian Space Agency, via del Politecnico snc, 00133 Roma, Italy\label{aff42}
\and
Dipartimento di Fisica, Universit\`a di Genova, Via Dodecaneso 33, 16146, Genova, Italy\label{aff43}
\and
INFN-Sezione di Genova, Via Dodecaneso 33, 16146, Genova, Italy\label{aff44}
\and
Department of Physics "E. Pancini", University Federico II, Via Cinthia 6, 80126, Napoli, Italy\label{aff45}
\and
INAF-Osservatorio Astronomico di Capodimonte, Via Moiariello 16, 80131 Napoli, Italy\label{aff46}
\and
Instituto de Astrof\'isica e Ci\^encias do Espa\c{c}o, Universidade do Porto, CAUP, Rua das Estrelas, PT4150-762 Porto, Portugal\label{aff47}
\and
Faculdade de Ci\^encias da Universidade do Porto, Rua do Campo de Alegre, 4150-007 Porto, Portugal\label{aff48}
\and
Dipartimento di Fisica, Universit\`a degli Studi di Torino, Via P. Giuria 1, 10125 Torino, Italy\label{aff49}
\and
INFN-Sezione di Torino, Via P. Giuria 1, 10125 Torino, Italy\label{aff50}
\and
INAF-Osservatorio Astrofisico di Torino, Via Osservatorio 20, 10025 Pino Torinese (TO), Italy\label{aff51}
\and
INAF-IASF Milano, Via Alfonso Corti 12, 20133 Milano, Italy\label{aff52}
\and
Centro de Investigaciones Energ\'eticas, Medioambientales y Tecnol\'ogicas (CIEMAT), Avenida Complutense 40, 28040 Madrid, Spain\label{aff53}
\and
Port d'Informaci\'{o} Cient\'{i}fica, Campus UAB, C. Albareda s/n, 08193 Bellaterra (Barcelona), Spain\label{aff54}
\and
Institute for Theoretical Particle Physics and Cosmology (TTK), RWTH Aachen University, 52056 Aachen, Germany\label{aff55}
\and
INAF-Osservatorio Astronomico di Roma, Via Frascati 33, 00078 Monteporzio Catone, Italy\label{aff56}
\and
INFN section of Naples, Via Cinthia 6, 80126, Napoli, Italy\label{aff57}
\and
Institute for Astronomy, University of Hawaii, 2680 Woodlawn Drive, Honolulu, HI 96822, USA\label{aff58}
\and
Dipartimento di Fisica e Astronomia "Augusto Righi" - Alma Mater Studiorum Universit\`a di Bologna, Viale Berti Pichat 6/2, 40127 Bologna, Italy\label{aff59}
\and
Institute for Astronomy, University of Edinburgh, Royal Observatory, Blackford Hill, Edinburgh EH9 3HJ, UK\label{aff60}
\and
European Space Agency/ESRIN, Largo Galileo Galilei 1, 00044 Frascati, Roma, Italy\label{aff61}
\and
Universit\'e Claude Bernard Lyon 1, CNRS/IN2P3, IP2I Lyon, UMR 5822, Villeurbanne, F-69100, France\label{aff62}
\and
Institut de Ci\`{e}ncies del Cosmos (ICCUB), Universitat de Barcelona (IEEC-UB), Mart\'{i} i Franqu\`{e}s 1, 08028 Barcelona, Spain\label{aff63}
\and
Instituci\'o Catalana de Recerca i Estudis Avan\c{c}ats (ICREA), Passeig de Llu\'{\i}s Companys 23, 08010 Barcelona, Spain\label{aff64}
\and
UCB Lyon 1, CNRS/IN2P3, IUF, IP2I Lyon, 4 rue Enrico Fermi, 69622 Villeurbanne, France\label{aff65}
\and
Mullard Space Science Laboratory, University College London, Holmbury St Mary, Dorking, Surrey RH5 6NT, UK\label{aff66}
\and
Departamento de F\'isica, Faculdade de Ci\^encias, Universidade de Lisboa, Edif\'icio C8, Campo Grande, PT1749-016 Lisboa, Portugal\label{aff67}
\and
Instituto de Astrof\'isica e Ci\^encias do Espa\c{c}o, Faculdade de Ci\^encias, Universidade de Lisboa, Campo Grande, 1749-016 Lisboa, Portugal\label{aff68}
\and
Department of Astronomy, University of Geneva, ch. d'Ecogia 16, 1290 Versoix, Switzerland\label{aff69}
\and
INAF-Istituto di Astrofisica e Planetologia Spaziali, via del Fosso del Cavaliere, 100, 00100 Roma, Italy\label{aff70}
\and
INFN-Padova, Via Marzolo 8, 35131 Padova, Italy\label{aff71}
\and
Aix-Marseille Universit\'e, CNRS/IN2P3, CPPM, Marseille, France\label{aff72}
\and
Max Planck Institute for Extraterrestrial Physics, Giessenbachstr. 1, 85748 Garching, Germany\label{aff73}
\and
Universit\"ats-Sternwarte M\"unchen, Fakult\"at f\"ur Physik, Ludwig-Maximilians-Universit\"at M\"unchen, Scheinerstrasse 1, 81679 M\"unchen, Germany\label{aff74}
\and
INFN-Bologna, Via Irnerio 46, 40126 Bologna, Italy\label{aff75}
\and
School of Physics, HH Wills Physics Laboratory, University of Bristol, Tyndall Avenue, Bristol, BS8 1TL, UK\label{aff76}
\and
NRC Herzberg, 5071 West Saanich Rd, Victoria, BC V9E 2E7, Canada\label{aff77}
\and
Institute of Theoretical Astrophysics, University of Oslo, P.O. Box 1029 Blindern, 0315 Oslo, Norway\label{aff78}
\and
Leiden Observatory, Leiden University, Einsteinweg 55, 2333 CC Leiden, The Netherlands\label{aff79}
\and
Jet Propulsion Laboratory, California Institute of Technology, 4800 Oak Grove Drive, Pasadena, CA, 91109, USA\label{aff80}
\and
Felix Hormuth Engineering, Goethestr. 17, 69181 Leimen, Germany\label{aff81}
\and
Technical University of Denmark, Elektrovej 327, 2800 Kgs. Lyngby, Denmark\label{aff82}
\and
Cosmic Dawn Center (DAWN), Denmark\label{aff83}
\and
Max-Planck-Institut f\"ur Astronomie, K\"onigstuhl 17, 69117 Heidelberg, Germany\label{aff84}
\and
NASA Goddard Space Flight Center, Greenbelt, MD 20771, USA\label{aff85}
\and
Department of Physics and Helsinki Institute of Physics, Gustaf H\"allstr\"omin katu 2, 00014 University of Helsinki, Finland\label{aff86}
\and
Universit\'e de Gen\`eve, D\'epartement de Physique Th\'eorique and Centre for Astroparticle Physics, 24 quai Ernest-Ansermet, CH-1211 Gen\`eve 4, Switzerland\label{aff87}
\and
Department of Physics, P.O. Box 64, 00014 University of Helsinki, Finland\label{aff88}
\and
Helsinki Institute of Physics, Gustaf H{\"a}llstr{\"o}min katu 2, University of Helsinki, Helsinki, Finland\label{aff89}
\and
Centre de Calcul de l'IN2P3/CNRS, 21 avenue Pierre de Coubertin 69627 Villeurbanne Cedex, France\label{aff90}
\and
Laboratoire d'etude de l'Univers et des phenomenes eXtremes, Observatoire de Paris, Universit\'e PSL, Sorbonne Universit\'e, CNRS, 92190 Meudon, France\label{aff91}
\and
Aix-Marseille Universit\'e, CNRS, CNES, LAM, Marseille, France\label{aff92}
\and
NOVA optical infrared instrumentation group at ASTRON, Oude Hoogeveensedijk 4, 7991PD, Dwingeloo, The Netherlands\label{aff93}
\and
Dipartimento di Fisica "Aldo Pontremoli", Universit\`a degli Studi di Milano, Via Celoria 16, 20133 Milano, Italy\label{aff94}
\and
INFN-Sezione di Milano, Via Celoria 16, 20133 Milano, Italy\label{aff95}
\and
University of Applied Sciences and Arts of Northwestern Switzerland, School of Computer Science, 5210 Windisch, Switzerland\label{aff96}
\and
Universit\"at Bonn, Argelander-Institut f\"ur Astronomie, Auf dem H\"ugel 71, 53121 Bonn, Germany\label{aff97}
\and
INFN-Sezione di Roma, Piazzale Aldo Moro, 2 - c/o Dipartimento di Fisica, Edificio G. Marconi, 00185 Roma, Italy\label{aff98}
\and
Dipartimento di Fisica e Astronomia "Augusto Righi" - Alma Mater Studiorum Universit\`a di Bologna, via Piero Gobetti 93/2, 40129 Bologna, Italy\label{aff99}
\and
Department of Physics, Institute for Computational Cosmology, Durham University, South Road, Durham, DH1 3LE, UK\label{aff100}
\and
University of Applied Sciences and Arts of Northwestern Switzerland, School of Engineering, 5210 Windisch, Switzerland\label{aff101}
\and
Institut d'Astrophysique de Paris, 98bis Boulevard Arago, 75014, Paris, France\label{aff102}
\and
Institute of Physics, Laboratory of Astrophysics, Ecole Polytechnique F\'ed\'erale de Lausanne (EPFL), Observatoire de Sauverny, 1290 Versoix, Switzerland\label{aff103}
\and
Aurora Technology for European Space Agency (ESA), Camino bajo del Castillo, s/n, Urbanizacion Villafranca del Castillo, Villanueva de la Ca\~nada, 28692 Madrid, Spain\label{aff104}
\and
Institut de F\'{i}sica d'Altes Energies (IFAE), The Barcelona Institute of Science and Technology, Campus UAB, 08193 Bellaterra (Barcelona), Spain\label{aff105}
\and
European Space Agency/ESTEC, Keplerlaan 1, 2201 AZ Noordwijk, The Netherlands\label{aff106}
\and
School of Mathematics, Statistics and Physics, Newcastle University, Herschel Building, Newcastle-upon-Tyne, NE1 7RU, UK\label{aff107}
\and
DARK, Niels Bohr Institute, University of Copenhagen, Jagtvej 155, 2200 Copenhagen, Denmark\label{aff108}
\and
Waterloo Centre for Astrophysics, University of Waterloo, Waterloo, Ontario N2L 3G1, Canada\label{aff109}
\and
Department of Physics and Astronomy, University of Waterloo, Waterloo, Ontario N2L 3G1, Canada\label{aff110}
\and
Perimeter Institute for Theoretical Physics, Waterloo, Ontario N2L 2Y5, Canada\label{aff111}
\and
Universit\'e Paris-Saclay, Universit\'e Paris Cit\'e, CEA, CNRS, AIM, 91191, Gif-sur-Yvette, France\label{aff112}
\and
Institute of Space Science, Str. Atomistilor, nr. 409 M\u{a}gurele, Ilfov, 077125, Romania\label{aff113}
\and
Dipartimento di Fisica e Astronomia "G. Galilei", Universit\`a di Padova, Via Marzolo 8, 35131 Padova, Italy\label{aff114}
\and
Institut f\"ur Theoretische Physik, University of Heidelberg, Philosophenweg 16, 69120 Heidelberg, Germany\label{aff115}
\and
Institut de Recherche en Astrophysique et Plan\'etologie (IRAP), Universit\'e de Toulouse, CNRS, UPS, CNES, 14 Av. Edouard Belin, 31400 Toulouse, France\label{aff116}
\and
Universit\'e St Joseph; Faculty of Sciences, Beirut, Lebanon\label{aff117}
\and
Departamento de F\'isica, FCFM, Universidad de Chile, Blanco Encalada 2008, Santiago, Chile\label{aff118}
\and
Institut d'Estudis Espacials de Catalunya (IEEC),  Edifici RDIT, Campus UPC, 08860 Castelldefels, Barcelona, Spain\label{aff119}
\and
Satlantis, University Science Park, Sede Bld 48940, Leioa-Bilbao, Spain\label{aff120}
\and
Instituto de Astrof\'isica e Ci\^encias do Espa\c{c}o, Faculdade de Ci\^encias, Universidade de Lisboa, Tapada da Ajuda, 1349-018 Lisboa, Portugal\label{aff121}
\and
Cosmic Dawn Center (DAWN)\label{aff122}
\and
Niels Bohr Institute, University of Copenhagen, Jagtvej 128, 2200 Copenhagen, Denmark\label{aff123}
\and
Universidad Polit\'ecnica de Cartagena, Departamento de Electr\'onica y Tecnolog\'ia de Computadoras,  Plaza del Hospital 1, 30202 Cartagena, Spain\label{aff124}
\and
Infrared Processing and Analysis Center, California Institute of Technology, Pasadena, CA 91125, USA\label{aff125}
\and
Dipartimento di Fisica e Scienze della Terra, Universit\`a degli Studi di Ferrara, Via Giuseppe Saragat 1, 44122 Ferrara, Italy\label{aff126}
\and
Istituto Nazionale di Fisica Nucleare, Sezione di Ferrara, Via Giuseppe Saragat 1, 44122 Ferrara, Italy\label{aff127}
\and
INAF, Istituto di Radioastronomia, Via Piero Gobetti 101, 40129 Bologna, Italy\label{aff128}
\and
Universit\'e C\^{o}te d'Azur, Observatoire de la C\^{o}te d'Azur, CNRS, Laboratoire Lagrange, Bd de l'Observatoire, CS 34229, 06304 Nice cedex 4, France\label{aff129}
\and
INAF-Osservatorio Astronomico di Brera, Via Brera 28, 20122 Milano, Italy, and INFN-Sezione di Genova, Via Dodecaneso 33, 16146, Genova, Italy\label{aff130}
\and
ICL, Junia, Universit\'e Catholique de Lille, LITL, 59000 Lille, France\label{aff131}
\and
ICSC - Centro Nazionale di Ricerca in High Performance Computing, Big Data e Quantum Computing, Via Magnanelli 2, Bologna, Italy\label{aff132}
\and
Instituto de F\'isica Te\'orica UAM-CSIC, Campus de Cantoblanco, 28049 Madrid, Spain\label{aff133}
\and
CERCA/ISO, Department of Physics, Case Western Reserve University, 10900 Euclid Avenue, Cleveland, OH 44106, USA\label{aff134}
\and
Technical University of Munich, TUM School of Natural Sciences, Physics Department, James-Franck-Str.~1, 85748 Garching, Germany\label{aff135}
\and
Max-Planck-Institut f\"ur Astrophysik, Karl-Schwarzschild-Str.~1, 85748 Garching, Germany\label{aff136}
\and
Laboratoire Univers et Th\'eorie, Observatoire de Paris, Universit\'e PSL, Universit\'e Paris Cit\'e, CNRS, 92190 Meudon, France\label{aff137}
\and
Departamento de F{\'\i}sica Fundamental. Universidad de Salamanca. Plaza de la Merced s/n. 37008 Salamanca, Spain\label{aff138}
\and
Universit\'e de Strasbourg, CNRS, Observatoire astronomique de Strasbourg, UMR 7550, 67000 Strasbourg, France\label{aff139}
\and
Center for Data-Driven Discovery, Kavli IPMU (WPI), UTIAS, The University of Tokyo, Kashiwa, Chiba 277-8583, Japan\label{aff140}
\and
Ludwig-Maximilians-University, Schellingstrasse 4, 80799 Munich, Germany\label{aff141}
\and
Max-Planck-Institut f\"ur Physik, Boltzmannstr. 8, 85748 Garching, Germany\label{aff142}
\and
Dipartimento di Fisica - Sezione di Astronomia, Universit\`a di Trieste, Via Tiepolo 11, 34131 Trieste, Italy\label{aff143}
\and
California Institute of Technology, 1200 E California Blvd, Pasadena, CA 91125, USA\label{aff144}
\and
Department of Physics \& Astronomy, University of California Irvine, Irvine CA 92697, USA\label{aff145}
\and
Department of Mathematics and Physics E. De Giorgi, University of Salento, Via per Arnesano, CP-I93, 73100, Lecce, Italy\label{aff146}
\and
INFN, Sezione di Lecce, Via per Arnesano, CP-193, 73100, Lecce, Italy\label{aff147}
\and
INAF-Sezione di Lecce, c/o Dipartimento Matematica e Fisica, Via per Arnesano, 73100, Lecce, Italy\label{aff148}
\and
Departamento F\'isica Aplicada, Universidad Polit\'ecnica de Cartagena, Campus Muralla del Mar, 30202 Cartagena, Murcia, Spain\label{aff149}
\and
Universit\'e Paris Cit\'e, CNRS, Astroparticule et Cosmologie, 75013 Paris, France\label{aff150}
\and
CEA Saclay, DFR/IRFU, Service d'Astrophysique, Bat. 709, 91191 Gif-sur-Yvette, France\label{aff151}
\and
Institute of Cosmology and Gravitation, University of Portsmouth, Portsmouth PO1 3FX, UK\label{aff152}
\and
Department of Computer Science, Aalto University, PO Box 15400, Espoo, FI-00 076, Finland\label{aff153}
\and
Instituto de Astrof\'\i sica de Canarias, c/ Via Lactea s/n, La Laguna 38200, Spain. Departamento de Astrof\'\i sica de la Universidad de La Laguna, Avda. Francisco Sanchez, La Laguna, 38200, Spain\label{aff154}
\and
Caltech/IPAC, 1200 E. California Blvd., Pasadena, CA 91125, USA\label{aff155}
\and
Ruhr University Bochum, Faculty of Physics and Astronomy, Astronomical Institute (AIRUB), German Centre for Cosmological Lensing (GCCL), 44780 Bochum, Germany\label{aff156}
\and
Department of Physics and Astronomy, Vesilinnantie 5, 20014 University of Turku, Finland\label{aff157}
\and
Serco for European Space Agency (ESA), Camino bajo del Castillo, s/n, Urbanizacion Villafranca del Castillo, Villanueva de la Ca\~nada, 28692 Madrid, Spain\label{aff158}
\and
ARC Centre of Excellence for Dark Matter Particle Physics, Melbourne, Australia\label{aff159}
\and
Centre for Astrophysics \& Supercomputing, Swinburne University of Technology,  Hawthorn, Victoria 3122, Australia\label{aff160}
\and
Department of Physics and Astronomy, University of the Western Cape, Bellville, Cape Town, 7535, South Africa\label{aff161}
\and
DAMTP, Centre for Mathematical Sciences, Wilberforce Road, Cambridge CB3 0WA, UK\label{aff162}
\and
Kavli Institute for Cosmology Cambridge, Madingley Road, Cambridge, CB3 0HA, UK\label{aff163}
\and
Department of Astrophysics, University of Zurich, Winterthurerstrasse 190, 8057 Zurich, Switzerland\label{aff164}
\and
Department of Physics, Centre for Extragalactic Astronomy, Durham University, South Road, Durham, DH1 3LE, UK\label{aff165}
\and
IRFU, CEA, Universit\'e Paris-Saclay 91191 Gif-sur-Yvette Cedex, France\label{aff166}
\and
Oskar Klein Centre for Cosmoparticle Physics, Department of Physics, Stockholm University, Stockholm, SE-106 91, Sweden\label{aff167}
\and
Astrophysics Group, Blackett Laboratory, Imperial College London, London SW7 2AZ, UK\label{aff168}
\and
Univ. Grenoble Alpes, CNRS, Grenoble INP, LPSC-IN2P3, 53, Avenue des Martyrs, 38000, Grenoble, France\label{aff169}
\and
INAF-Osservatorio Astrofisico di Arcetri, Largo E. Fermi 5, 50125, Firenze, Italy\label{aff170}
\and
Dipartimento di Fisica, Sapienza Universit\`a di Roma, Piazzale Aldo Moro 2, 00185 Roma, Italy\label{aff171}
\and
Centro de Astrof\'{\i}sica da Universidade do Porto, Rua das Estrelas, 4150-762 Porto, Portugal\label{aff172}
\and
HE Space for European Space Agency (ESA), Camino bajo del Castillo, s/n, Urbanizacion Villafranca del Castillo, Villanueva de la Ca\~nada, 28692 Madrid, Spain\label{aff173}
\and
Department of Astrophysical Sciences, Peyton Hall, Princeton University, Princeton, NJ 08544, USA\label{aff174}
\and
Theoretical astrophysics, Department of Physics and Astronomy, Uppsala University, Box 515, 751 20 Uppsala, Sweden\label{aff175}
\and
Mathematical Institute, University of Leiden, Einsteinweg 55, 2333 CA Leiden, The Netherlands\label{aff176}
\and
Institute of Astronomy, University of Cambridge, Madingley Road, Cambridge CB3 0HA, UK\label{aff177}
\and
Space physics and astronomy research unit, University of Oulu, Pentti Kaiteran katu 1, FI-90014 Oulu, Finland\label{aff178}
\and
Center for Computational Astrophysics, Flatiron Institute, 162 5th Avenue, 10010, New York, NY, USA\label{aff179}
\and
Department of Physics and Astronomy, University of British Columbia, Vancouver, BC V6T 1Z1, Canada\label{aff180}}

   \date{Received September 15, 1996; accepted March 16, 1997}


\abstract
{ Stellar bars are key structures in disc galaxies, driving angular momentum redistribution and influencing processes such as bulge growth and star formation. Quantifying the bar fraction as a function of redshift and stellar mass is therefore important for constraining the physical processes that drive disc formation and evolution across the history of the Universe. Leveraging the unprecedented resolution and survey area of the \Euclid{} Q1 data release combined with the \texttt{Zoobot} deep-learning model trained on citizen-science labels, we identify \(7711\) barred galaxies with \(M_* \gtrsim 10^{10}\,M_\odot\) in a magnitude-selected sample \((I_E < 20.5)\) spanning \(63.1\,\mathrm{deg}^2\). We measure a mean bar fraction of \(0.2-0.4\), consistent with prior studies. At fixed redshift, massive galaxies exhibit higher bar fractions, while lower-mass systems show a steeper decline with redshift, suggesting earlier disc assembly in massive galaxies. Comparisons with cosmological simulations (e.g., \texttt{TNG50}, \texttt{Auriga}) reveal a broadly consistent bar fraction, but highlight overpredictions for high-mass systems, pointing to potential over-efficiency in central stellar mass build-up in simulations. These findings demonstrate \Euclid{}’s transformative potential for galaxy morphology studies and underscore the importance of refining theoretical models to better reproduce observed trends. Future work will explore finer mass bins, environmental correlations, and additional morphological indicators.}

   \keywords{Galaxies: evolution - Galaxies: fundamental parameters - Galaxies: high-redshift
               }

   \maketitle
%
\section{Introduction}
Stellar bars, which are elongated stellar structures extending from the central regions of disc galaxies, represent a fundamental dynamical component of galaxies. They play a critical role in redistributing angular momentum within galaxies, driving secular evolution processes such as central bulge growth, fuelling active galactic nuclei (AGN), and triggering episodes of star formation \citep[e.g.,][]{2003MNRAS.341.1179A,2004ARA&A..42..603K}. 

The formation of bars is primarily governed by disc instabilities. Classical theoretical studies and simulations suggest that bars can form naturally in dynamically cold discs over timescales of a few gigayears, with their strength and longevity depending on factors such as the galaxy gas content, dark matter halo, and internal stellar velocity dispersion  \citep[e.g.,][]{2000ApJ...543..704D, 2003MNRAS.341.1179A}. However, the discovery of barred galaxies at very early epochs following the launch of the \textit{James Webb} Space Telescope (JWST) has triggered new interest on the physical mechanisms responsible for bar formation (e.g.,~\citealp{2023A&A...678A..54M,2023Natur.623..499C,2024arXiv240906100G, 2024MNRAS.530.1984L}). The high gas fractions and turbulent conditions of the early Universe disfavour bar formation according to the classical view, which is supported by some observational evidence from the local Universe~\citep{2012MNRAS.424.2180M}.
Recent simulations suggest in fact that the ratio between dark matter and baryonic matter might play a key role in regulating bar formation (e.g.,~\citealp{2018MNRAS.477.1451F,2022MNRAS.512..160R,2024MNRAS.529..979L,  2024arXiv240609453F}) .


Understanding the fraction of barred galaxies as a function of redshift and stellar mass thus provides valuable insights into the formation and growth of stellar discs across cosmic time and baryon assembly more generally (e.g., ~\citealp{2004ApJ...615L.105J, 2008ApJ...675.1141S, 2010MNRAS.409..346C, 2011MNRAS.411.2026M, 2014MNRAS.445.3466S, 2014MNRAS.438.2882M, 2018MNRAS.474.5372E,2024arXiv240906100G}). 

Identifying bars in galaxies typically requires high-resolution imaging to discern the distinct morphology of barred structures. Historically, visual classification has been a powerful tool for bar identification \citep[e.g.,][]{2000AJ....119..536E, 2011MNRAS.411.2026M, 2014MNRAS.445.3466S}, complemented by quantitative methods such as ellipse fitting \citep{2000ApJ...529...93K, 2009A&A...495..491A}, Fourier decomposition \citep{1990ApJ...357...71O}, and more recently machine learning \citep[e.g.,][]{2018MNRAS.476.3661D,2022MNRAS.509.3966W}. Previous studies using data from space facilities like the \textit{Hubble} Space Telescope (HST) and JWST have significantly advanced our understanding of barred galaxies, particularly beyond the local Universe \citep[e.g.,][]{2008ApJ...675.1141S, 2014MNRAS.438.2882M}. However, these studies are often constrained by a limited area coverage, which hinders a comprehensive statistical analysis across diverse galaxy populations.

The \textit{Euclid} space telescope represents a transformative step forward in this field. \textit{Euclid} combines high spatial resolution and sensitivity with an unprecedented survey area for a space based observatory, enabling a detailed study of galaxy morphology on a new scale \citep{Laureijs11, Scaramella-EP1,Bretonniere-EP13,Bretonniere-EP26,EuclidSkyOverview, EP-Aussel}. The \textit{Euclid} Q1 data release which provides high-quality imaging over 63.1 ${\rm deg}^2$~\citep{Q1cite,Q1-TP001}, already represents a dramatic increase of the area probed by previous space observatories such as HST and JWST. The largest optical HST survey, the Cosmic Evolution Survey \citep[COSMOS,][]{2007ApJS..172....1S}, covers an area of only about $2$ deg$^2$. 

In this work, we leverage the unique capabilities of \textit{Euclid} to provide a first measurement of the fraction of barred galaxies in massive systems (stellar masses $M_* \gtrsim 10^{10} M_\odot$) up to redshift $z \sim 1$ using deep-learning classifications trained on visual inspections. This work increases the number of barred galaxies by more than an order of magnitude compared to prior studies based on HST and JWST data, providing a robust reference of the abundance of bars in massive galaxies over half of cosmic history.

The paper proceeds as follows. Section~\ref{sec:data} describes the data used for this work, namely the \Euclid Q1 data release. Section~\ref{sec:bars} details the procedure employed to select bars. The main results  are an exploration of the evolution of the bar fraction as a function of stellar mass and redshift. These results are explored in Sect.~\ref{sec:results} and discussed in Sect.~\ref{sec:disc}, where we compare with previous observational and simulated results.

\section{Data and measurements}
\label{sec:data}
\subsection{\Euclid Q1 data release}

This work uses data from the \Euclid Q1 data release~\citep{Q1-TP001}. An extended description of the \Euclid mission and scientific objectives can be found in~\cite{EuclidSkyOverview}. The Q1 data release comprises an area of $63.1$ $\deg^2$ distributed in three distinct fields: Euclid Deep Field North (EDF-N); Euclid Deep Field South (EDF-S); and Euclid Deep Field Fornax (EDF-F). All fields are observed with both the VIS~\citep{EuclidSkyVIS} and NISP~\citep{EuclidSkyNISP} instruments. A detailed description of the Q1 data release is presented in~\cite{Q1-TP001} and specific details about the VIS and NISP data products can be consulted in~\cite{Q1-TP002} and~\cite{Q1-TP003}, respectively. For this particular work, we employ a number of data products accompanying the data release, accessible from the \Euclid Science Archive System (SAS) which we detail in the following.  

\subsubsection{\Euclid Q1 detailed morphology catalogue}
\label{sec:catalog}

The Q1 data release contains a variety of morphological measurements for detected galaxies, including non-parametric morphologies, parametric S\'ersic fits, and deep learning-based detailed visual like morphologies. We refer the reader to \cite{Q1-TP004} for an extensive description of the \Euclid photometric catalogue. 

For this work, we make primarily use of the detailed morphological catalogue (see~\citealp{Q1-SP047} for more details). In a nutshell, the catalogue contains Galaxy Zoo (GZ) type classifications, following the tree structure of the GZ-CANDELS project \citep{2017MNRAS.464.4420S}, which uses data from the Cosmic Assembly Near-infrared Deep Extragalactic Legacy Survey (CANDELS). The classifications have been performed using the \texttt{Zoobot} deep-foundation model~\citep{2022zndo...6483176W}. The model has been fine tuned with volunteer classifications of \Euclid galaxies obtained between August and September 2024. As detailed in~\cite{Q1-SP047}, three different images were shown to the GZ volunteers to label the galaxies: an RGB image where the R channel is \YE, the B channel is \IE, and the G channel is the mean, following a clip and an arcsinh stretch; a greyscale image where the single channel is the same as the \IE/B channel of the RGB image for maximising resolution; and a greyscale image where the single channel is again from \IE, but adjusted to highlight low surface brightness features in the outskirts of the galaxies. A complete description of the data product as well as a quantitative assessment of the accuracy is presented in the accompanying work~\citep{Q1-SP047}. In Sect.~\ref{sec:bars} we describe in more detail the procedure employed for selecting bars.

\subsubsection{\Euclid Q1 physical properties}
In addition to morphologies we use photometric redshifts and stellar masses from the data release. More details can be found in~\cite{Q1-TP005}. Briefly, a large grid of synthetic galaxy spectral energy distribution (SED) models is generated using the \texttt{Bagpipes} package \citep{2018MNRAS.480.4379C} with delayed exponential star-formation histories. These models are fit to the Q1 galaxies with the software \texttt{NNPZ}~\citep{Q1-TP005}, whereby the closest 30 models in $\chi^2$ are used to form a posterior distribution of the galaxy physical properties. In this work we use the marginalised medians of the posterior as our point estimate in redshift and stellar mass. 


\subsection{Sample selection and completeness}

\begin{figure}[t]
  \centering
  \setlength{\abovecaptionskip}{-3mm}
    \includegraphics[width=1\columnwidth]{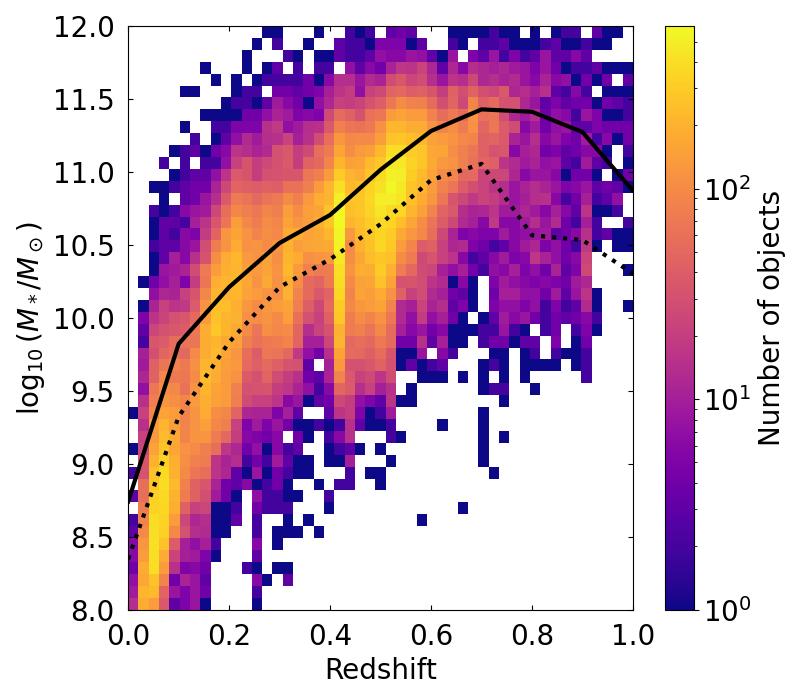}
 \caption{Photo-$z$ vs. stellar mass diagram showing the completeness limits for the Euclid Q1-GZ data set. The $90\%$ and $50\%$ stellar mass completeness limits are derived following~\cite{2010A&A...523A..13P} and are indicated by the solid and dotted black lines respectively. 
 }
  \label{fig:photo-z_smass}
\end{figure}

The \Euclid Q1 deep learning morphological classification~\citep{Q1-TP004, Q1-SP047} is provided only for galaxies with $\IE<20.5$ or with a segmentation area larger than $1200$ pixels. Although \Euclid data allow us to measure accurate morphologies for fainter and smaller galaxies~\citep{Bretonniere-EP13}, these conservative cuts have been selected to ensure very robust morphologies for this first data release~\citep{EP-Aussel}. Therefore, for the remainder of this work, we only use galaxies brighter than $\IE=20.5$. This stringent selection severely impacts the completeness of the sample, which needs to be carefully addressed before deriving any scientific conclusion. Figure~\ref{fig:photo-z_smass} shows the photometric redshift -- stellar mass plane. We compute a $90\%$ stellar mass completeness using the method from~\cite{2010A&A...523A..13P}. We find that the stellar mass above which the sample is $90\%$ complete rapidly increases with redshift, being around $10^{11}M_\odot$ at $z>0.6$. This is a direct consequence of the very bright  magnitude cut applied. To keep enough statistics while limiting the impact of incompleteness, we keep galaxies with stellar masses larger than $10^{10}M_\odot$ in the analysis. However, since this stellar mass threshold is significantly below the completeness limit, especially at high redshift, we adopt narrow stellar mass bins for analysing evolutionary trends and discuss the impact of this choice on the results of this work. 

\section{Bar classification }
\label{sec:bars}
\subsection{\texttt{Zoobot} classifications}
The main result of this work is a derivation of the evolution of the bar fraction up to $z\sim1$. Barred galaxies are selected using the \texttt{Zoobot} classifications included in the \Euclid MER morphological catalogue (see~\citealp{Q1-SP047} for more details).  \texttt{Zoobot} is a probabilistic deep-learning model trained to reproduce the GZ classification tree. To that purpose images are preprocessed following the GZ standard. This includes scaling to limit the impact of the large dynamic range as well as resizing so that all galaxies present a similar apparent size in the image. The model hence estimates for each galaxy the fraction of volunteers who would have been selected a given morphological feature, had this galaxy been classified by GZ. Full details of the preprocessing, model used, and the specific training strategy followed for \Euclid data can be found in~\cite{Q1-SP047}. 

Given the tree-like structure of the \texttt{Zoobot} classification, we apply the following criteria to select bars:
$$p_\textrm{feature}>0.5; p_\textrm{edge-on}<0.5;p_\textrm{bar}>0.5$$,
where $p_\textrm{feature},p_\textrm{edge-on},$ and $p_\textrm{bar}$ are the outputs of the \texttt{Zoobot} classification, measuring the fraction of votes for a galaxy to be classified as featured, edge-on, and hosting a bar, respectively. The first cut selects galaxies with resolved features as opposed to smooth galaxies for which the question on bars is not asked. The second cut removes edge-on discs for which identifying stellar bars is difficult. Given that this is a pure random projection effect, it should not induce any bias. Finally, the last cut selects galaxies that likely host a bar. The exact threshold used can be changed, resulting in different values of purity and completeness~\citep{Q1-SP047}. 

\begin{figure*}[t]
  \centering \includegraphics[width=1\textwidth]{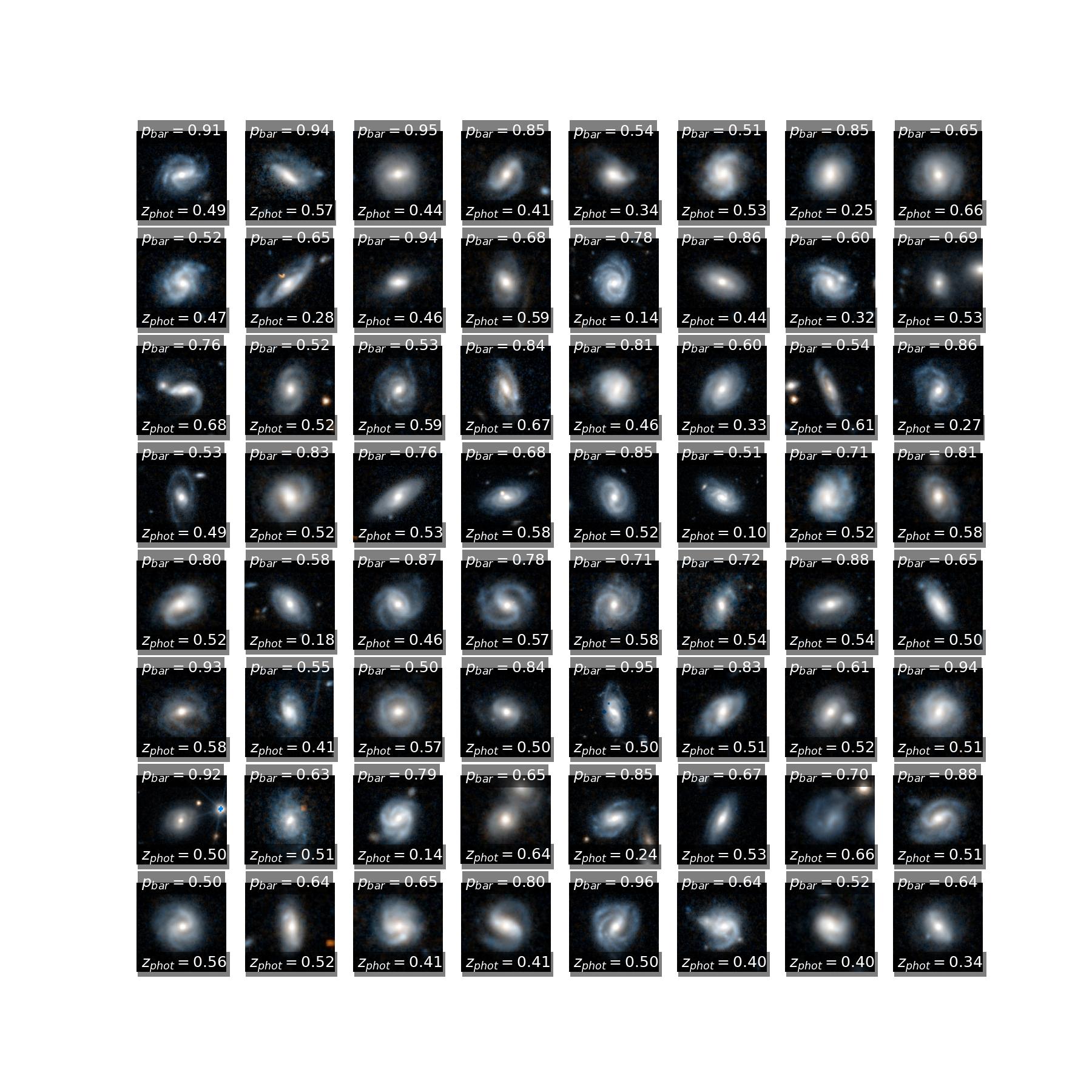}
  \vspace{-2.2cm}
 \caption{Random example colour cutouts of barred galaxies selected using the \Euclid Q1 morphology classification.  Most galaxies show a clear bar structure. The cutouts have been rescaled based on the effective radii of the galaxies so that they appear with a similar size to the volunteers (see~\citealp{Q1-SP047} for more details). }
  \label{fig:cutouts}
\end{figure*}

Figure~\ref{fig:cutouts} shows some random examples of barred galaxies selected using the criteria above. The vast majority shows a clear stellar bar, confirming the soundness of the classification. For a more detailed quantification we refer the reader to the accompanying work~\cite{Q1-SP047}.  The figure also suggests that the classification is mostly sensitive to strong bars. The impact of bar strength on the inferred bar fraction has triggered long debates over the past decades. It is well established that works based on GZ in the local Universe~\citep{2011MNRAS.411.2026M} tend to report a systematically smaller bar fraction (around $30\%$) than many morphological classifications on local samples with visual inspections done by professional astronomers~(e.g. \citealp{2000AJ....119..536E}), which report fractions larger than $60\%$. A first-order explanation for this discrepancy, put forward by~\cite{2011MNRAS.411.2026M}, is that the $30\%$ value found by GZ works mostly refers to strong bars and that weakly barred galaxies account for the difference with local studies.~\cite{2008ApJ...675.1141S} also showed that the fractions of barred galaxies at low redshift vary from about $60\%$ to $30\%$ if weak bars are 
 excluded from the sample. This, however, is not a fully settled story, since the concept of a strong bar is not very well defined in the literature. In addition,~\cite{2021MNRAS.507.4389G} showed that GZ classifications can be used to find weak bars with the proper selection. Another possibility is that GZ might trace prominent instead of strong bars~\citep{2018MNRAS.474.5372E} and hence fails to detect bars in low-mass, blue, and gas-rich galaxies; however, the appendix of~\cite{2018MNRAS.473.4731K} provides more evidence to link GZ bars with strong bars only. Finally, it is also known that the wavelength of observation has a significant impact on the sensitivity to identify bars. Emission from young stars, stronger in blue filters, and absorption by dust, tend to outshine or hide the presence of a stellar bar. This is why near infrared Observations are generally more suitable for exploring the abundance of bars~\citep{2000AJ....119..536E, 2000ApJ...529...93K}. Although the images used in this work are a composite of the \IE and \YE bands (Sect.~\ref{sec:catalog}), the higher spatial resolution of VIS likely dominates the classification. We will further discuss the impact of these limitations when discussing the results in Sect.~\ref{sec:disc}.


\subsection{Detection biases}
\label{sec:biases}
In addition to the \texttt{Zoobot} classification accuracy, which mimics the visual classification, it is important to quantify intrinsic biases due to signal-to-noise (S/N) and resolution differences. This is crucial for analysing redshift trends because classification biases
could falsely mimic such trends.  Galaxies at high redshift appear smaller and bar sizes are expected to evolve
with redshift. These factors could make it harder to detect bars, leading to an
apparent decrease in the bar fraction.



\begin{figure*}[t]
  \centering
  \setlength{\abovecaptionskip}{-3mm}
    \includegraphics[width=1\textwidth]{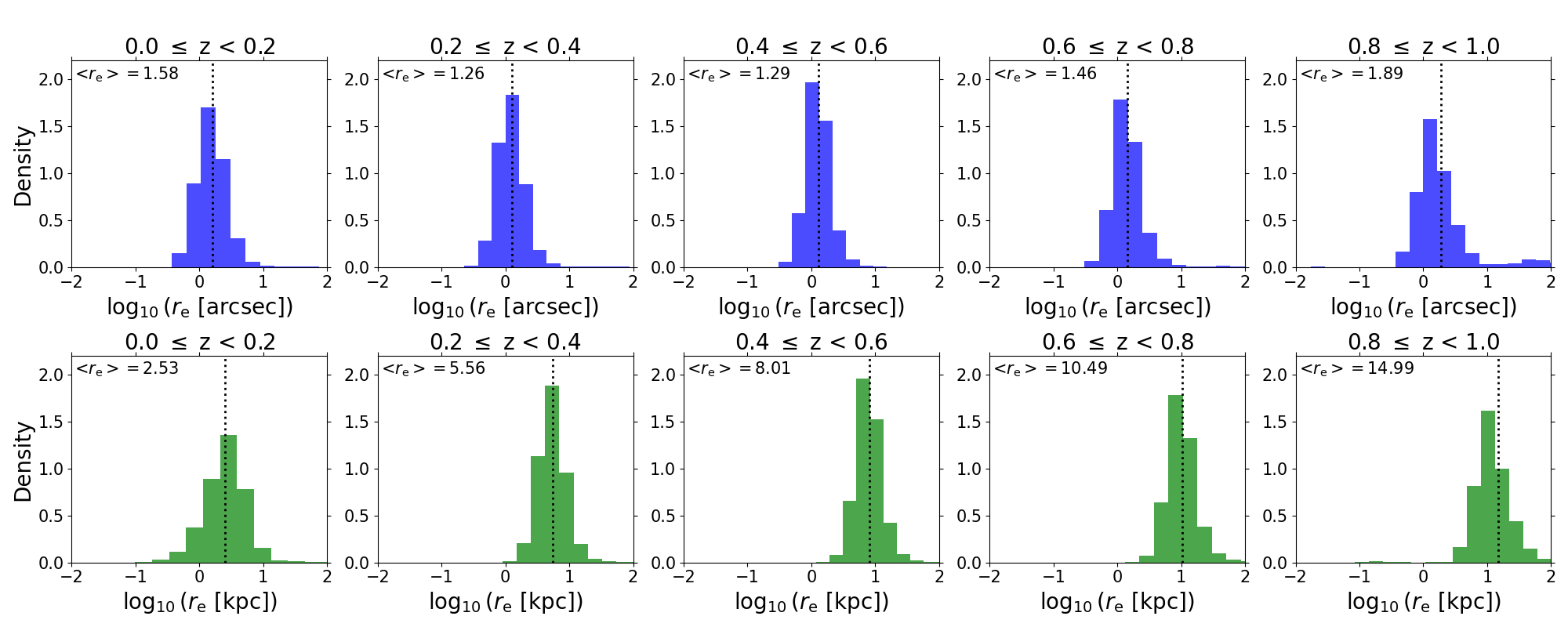}
 \caption{Distribution of apparent (top row) and physical sizes (bottom row) in different redshift bins as labelled. The vertical dashed lines indicate the mean values of each distribution which numerical value is also indicated in each panel. The bright magnitude cut applied implies a roughly constant apparent size with redshift. }
  \label{fig:size_histograms}
\end{figure*}

To quantify these effects, we first look in Fig.~\ref{fig:size_histograms} at the apparent and physical size distributions of galaxies in our sample as a function of redshift. Interestingly, the bright magnitude cut keeps the apparent effective radius ($r_{\rm e}$) of the sample
relatively constant with redshift ($r_{\rm e}\sim$1--2 arcsec). As long
as the relationship between bar length and effective radius~\citep{2019MNRAS.489.3553E} remains
stable, bar detection is unlikely to be significantly affected by declining
resolution at higher redshifts. Our selection implies that the galaxies that we analyse are on average intrinsically larger and more massive (Fig.~\ref{fig:photo-z_smass}) at higher redshifts and hence host larger bars compensating the degradation of the resolution. Figure~\ref{fig:size_histograms} indeed shows that the ratio $r_{\rm e}/\theta$ (where $\theta$ is the full-width half maximum of the point spread function) in physical units remains essentially constant in the redshift range explored.


\begin{figure}
\includegraphics[width=\columnwidth]{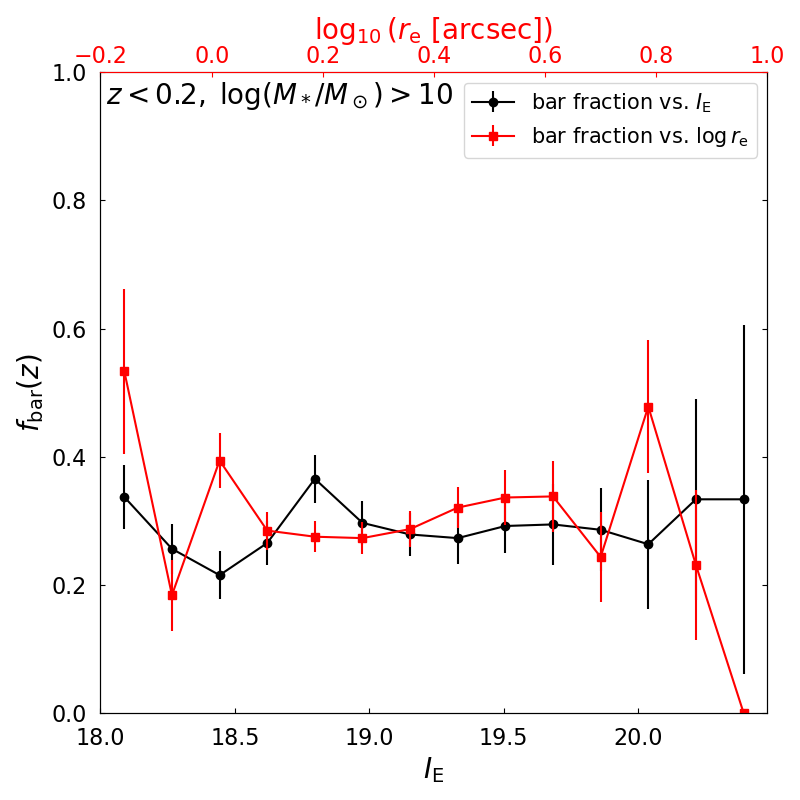}

 \caption{Detection bias of bars. The red and black solid lines show the bar fraction as a function of apparent effective radius (top $x$-axis) and apparent \IE magnitude (bottom $x$-axis), respectively, for galaxies at $z<0.2$. Error bars indicate the 68\% confidence interval under a beta-binomial posterior. The lack of trend suggests that the detection of bars is not affected by S/N and spatial resolution variations in the selected sample.
 }
  \label{fig:bar_frac_bias}
\end{figure}

Even though the size distributions are similar at different redshifts and galaxies are bright, there might be differences in the ability to detect bars between small and large galaxies and/or faint and bright galaxies in our sample, which can cause additional biases.  
We attempt to quantify the impact of S/N and spatial resolution in Fig.~\ref{fig:bar_frac_bias}. The figure shows the bar fraction (see Sect.~\ref{sec:results} for a formal definition) as a function of \IE magnitude and observed effective radii in a narrow bin of redshift ($z<0.2$).  Since we are exploring a narrow bin of stellar mass ($\logten (M_*/M_\odot)>10$) and redshift, one can assume that the bar fraction should not depend on apparent size or magnitude for an unbiased classification since we are looking at a subset of galaxies with similar physical properties. 
Figure~\ref{fig:bar_frac_bias} indeed shows almost no dependence of the bar fraction with \IE and $r_{\rm e}$, suggesting that the bar classification is unbiased for the conservative sample explored in this work. We hence do not apply any correction to the measured fraction of bars in the forthcoming analysis. However, it is important to emphasise that this comes at the expense of completeness, since we are only complete for very massive galaxies at $z>0.5$. 

\section{Results: evolution of the bar fraction at $z<1$}
\label{sec:results}

The bar fraction simply measures the frequency of barred galaxies ($N_{\rm bar}$) in a given population of galaxies ($N_{\rm gal}$):
$$f_{\rm bar}=\frac{N_{\rm bar}}{N_{\rm gal}}\textrm{.}$$
The numerator $N_{\rm bar}$ is computed using the selection criteria defined in Sect.~\ref{sec:bars}. Given the tree like structure of the \texttt{Zoobot} classifications, we define $N_\textrm{gal}$ as the number of featured  galaxies, excluding edge-on galaxies:

$$N_{\textrm{gal}} = \left| \{ p_{\textrm{feature}} > 0.5 \} \cap \{ p_{\textrm{edge-on}} < 0.5 \} \right|\textrm{.}$$

The first selection selects featured galaxies that according to the GZ classifications are objects with clearly defined internal structure as opposed to smooth galaxies. This separation is similar, but not identical, to a more traditional late-type/early-type classification. As noted in some previous work~\citep{2017MNRAS.464.4420S, 2022MNRAS.509.4024D}, some featureless discs can be classified as smooth. This is important when comparing the results on the bar fraction with previously published work in Sect.~\ref{sec:lit}.

\begin{figure*}[t]
  \centering
  \setlength{\abovecaptionskip}{-3mm}
     \includegraphics[width=\textwidth]{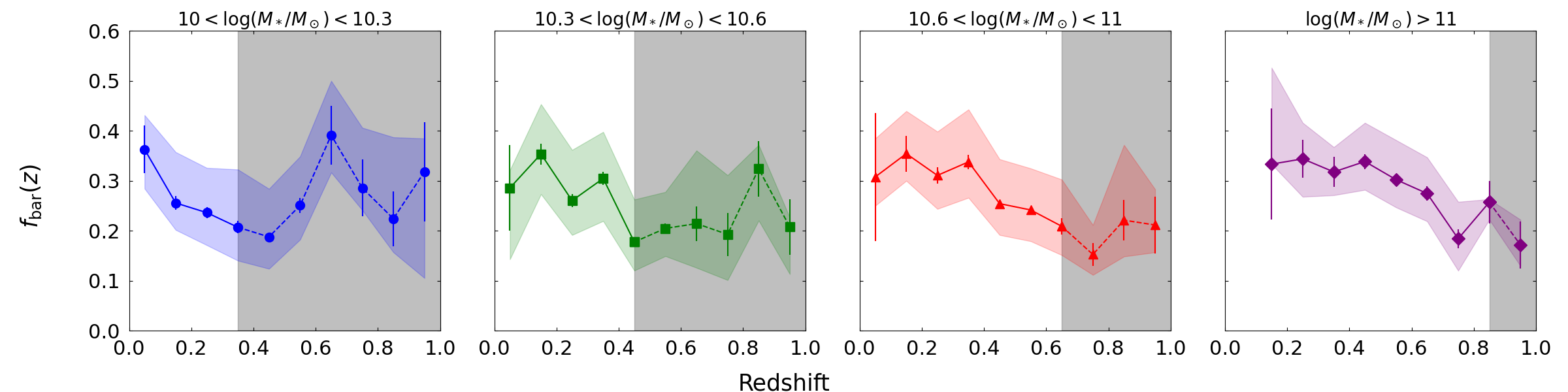}
 \caption{Evolution of the bar fraction as a function of redshift. Each panel shows a different stellar mass bin as labelled. The coloured shaded regions indicate the effect of changing the threshold for selecting barred galaxies between 0.4 and 0.6. The grey shaded regions indicate the redshift ranges affected by incompleteness. Error bars indicate the 68\% confidence interval under a beta-binomial posterior. The bar fraction shows a dependence with stellar mass, both in the normalisation and the evolutionary trends.  
 }
  \label{fig:euclid_bar_frac_relative}
\end{figure*}

Figure~\ref{fig:euclid_bar_frac_relative} shows the bar fraction as a function of redshift in bins of stellar mass. Table~\ref{tbl:nbars} reports the number of featured and barred galaxies, as well as the bar fraction in each redshift and stellar mass bin. Since the completeness of our sample strongly depends on stellar mass, we show the bar fraction for four different stellar mass bins and indicate the region of the parameter space where incompleteness starts to have a stronger impact based on the results of Fig.~\ref{fig:photo-z_smass}. We emphasise that Fig.~\ref{fig:euclid_bar_frac_relative} does not report a true evolution of the bar fraction along the progenitors, which should take into account the growth in mass. It rather shows variations of the bar fraction at fixed stellar mass. It is difficult to precisely quantify the effect of incompleteness in the measured bar fraction, since it depends on several unknowns such as the dependence of the bar fraction on effective radii and magnitude at fixed stellar mass and redshift. In addition, as previously mentioned, the exact normalisation of the bar fraction depends on a number of assumptions, such as the exact threshold to select barred galaxies or the denominator used. These systematic effects are particularly important because they dominate the error budget given the small statistical errors of \Euclid data. The shaded region in Fig.~\ref{fig:euclid_bar_frac_relative} indeed shows the impact of changing the \texttt{Zoobot} probability threshold to select barred galaxies from 0.4 to 0.6. It can change the bar fraction by $20\%$, although the main trends are preserved. 

Despite these known limitations, Fig.~\ref{fig:euclid_bar_frac_relative} reveals some interesting trends. We observe a moderate decrease of the bar fraction  with increasing redshift in all stellar mass bins. The decrease seems to be more pronounced at lower masses. In the low stellar mass bin ($10<\logten( M_*/M_\odot)<10.3$) the fraction drops from around $35\%$ to $20\%$ for $z\sim0$ to $z\sim0.3$. For the most massive galaxies ($\logten( M_*/M_\odot)>11$) the fraction remains almost constant over the same redshift range and only starts to noticeably decrease around $z\sim0.7$. Additionally, we also observe a slight dependence of the bar fraction with stellar mass at all redshifts. Massive galaxies present a slightly higher bar fraction than low mass galaxies at a similar redshift.





\begin{table}
  \centering
    \caption{Number of barred and featured galaxies in different redshift and stellar mass bins. The shaded rows indicate the areas of the parameter space most affected by incompleteness.}
  \begin{tabular}{llrrcl}
    \hline
    \hline
     & $z_\textrm{min}$--$z_\textrm{max}$ & $N_\textrm{featured}$ & $N_\textrm{bar}$ & $f_\textrm{bar}$ & \\ 
    \hline
    \noalign{\vskip 4pt}
    \multicolumn{6}{c}{\textbf{$10 < \logten(M_*/M_\odot) < 10.3$}} \\ 
    \hline
    \noalign{\vskip 4pt}
     & 0.00--0.10 & 102 & 37 & 0.363 & \\ 
     & 0.10--0.20 & 974 & 249 & 0.256 & \\ 
     & 0.20--0.30 & 1435 & 339 & 0.236 & \\ 
    \rowcolor[HTML]{D3D3D3}  & 0.30--0.40 & 1063 & 220 & 0.207 & \\ 
    \rowcolor[HTML]{D3D3D3}  & 0.40--0.50 & 2664 & 500 & 0.188 & \\ 
    \rowcolor[HTML]{D3D3D3}  & 0.50--0.60 & 802 & 201 & 0.251 & \\ 
    \rowcolor[HTML]{D3D3D3}  & 0.60--0.70 & 69 & 27 & 0.391 & \\ 
    \rowcolor[HTML]{D3D3D3}  & 0.70--0.80 & 63 & 18 & 0.286 & \\ 
    \rowcolor[HTML]{D3D3D3}  & 0.80--0.90 & 58 & 13 & 0.224 & \\ 
    \rowcolor[HTML]{D3D3D3}  & 0.90--1.00 & 22 & 7 & 0.318 & \\ 
    \hline
    \noalign{\vskip 4pt}
    \multicolumn{6}{c}{\textbf{$10.3 < \logten(M_*/M_\odot) < 10.6$}} \\ 
    \hline
    \noalign{\vskip 4pt}
     & 0.00--0.10 & 28 & 8 & 0.286 & \\ 
     & 0.10--0.20 & 509 & 180 & 0.354 & \\ 
     & 0.20--0.30 & 1214 & 316 & 0.260 & \\ 
     & 0.30--0.40 & 1267 & 386 & 0.305 & \\ 
     & 0.40--0.50 & 3369 & 598 & 0.178 & \\ 
    \rowcolor[HTML]{D3D3D3}  & 0.50--0.60 & 1729 & 354 & 0.205 & \\ 
    \rowcolor[HTML]{D3D3D3}  & 0.60--0.70 & 140 & 30 & 0.214 & \\ 
    \rowcolor[HTML]{D3D3D3}  & 0.70--0.80 & 83 & 16 & 0.193 & \\ 
    \rowcolor[HTML]{D3D3D3}  & 0.80--0.90 & 71 & 23 & 0.324 & \\ 
    \rowcolor[HTML]{D3D3D3}  & 0.90--1.00 & 53 & 11 & 0.208 & \\ 
    \hline
    \noalign{\vskip 4pt}
    \multicolumn{6}{c}{\textbf{$10.6 < \logten(M_*/M_\odot) < 11$}} \\ 
    \hline
    \noalign{\vskip 4pt}
     & 0.00--0.10 & 13 & 4 & 0.308 & \\ 
     & 0.10--0.20 & 175 & 62 & 0.354 & \\ 
     & 0.20--0.30 & 753 & 234 & 0.311 & \\ 
     & 0.30--0.40 & 1033 & 349 & 0.338 & \\ 
     & 0.40--0.50 & 3083 & 784 & 0.254 & \\ 
     & 0.50--0.60 & 3696 & 893 & 0.242 & \\ 
    \rowcolor[HTML]{D3D3D3}  & 0.60--0.70 & 608 & 127 & 0.209 & \\ 
    \rowcolor[HTML]{D3D3D3}  & 0.70--0.80 & 236 & 36 & 0.153 & \\ 
    \rowcolor[HTML]{D3D3D3}  & 0.80--0.90 & 104 & 23 & 0.221 & \\ 
    \rowcolor[HTML]{D3D3D3}  & 0.90--1.00 & 52 & 11 & 0.212 & \\ 
    \hline
    \noalign{\vskip 4pt}
    \multicolumn{6}{c}{\textbf{ $\logten(M_*/M_\odot) > 11$}} \\ 
    \hline
    \noalign{\vskip 4pt}
     & 0.00--0.10 & 2 & 0 & N/A & \\ 
     & 0.10--0.20 & 18 & 6 & 0.333 & \\ 
     & 0.20--0.30 & 157 & 54 & 0.344 & \\ 
     & 0.30--0.40 & 236 & 75 & 0.318 & \\ 
     & 0.40--0.50 & 1013 & 343 & 0.339 & \\ 
     & 0.50--0.60 & 2649 & 801 & 0.302 & \\ 
     & 0.60--0.70 & 954 & 262 & 0.275 & \\ 
     & 0.70--0.80 & 413 & 76 & 0.184 & \\ 
    \rowcolor[HTML]{D3D3D3}  & 0.80--0.90 & 105 & 27 & 0.257 & \\ 
    \rowcolor[HTML]{D3D3D3}  & 0.90--1.00 & 64 & 11 & 0.172 & \\ 
    \hline
  \end{tabular}

  \label{tbl:nbars}
\end{table}

\section{Discussion}
\label{sec:disc}

We now discuss the results presented in this paper in light of previous observational work and predictions from cosmological simulations.

\subsection{Comparison with previous observational results}
\label{sec:lit}

\begin{figure}[t]
  \centering
  \setlength{\abovecaptionskip}{-3mm} 
  \includegraphics[width=1\columnwidth]{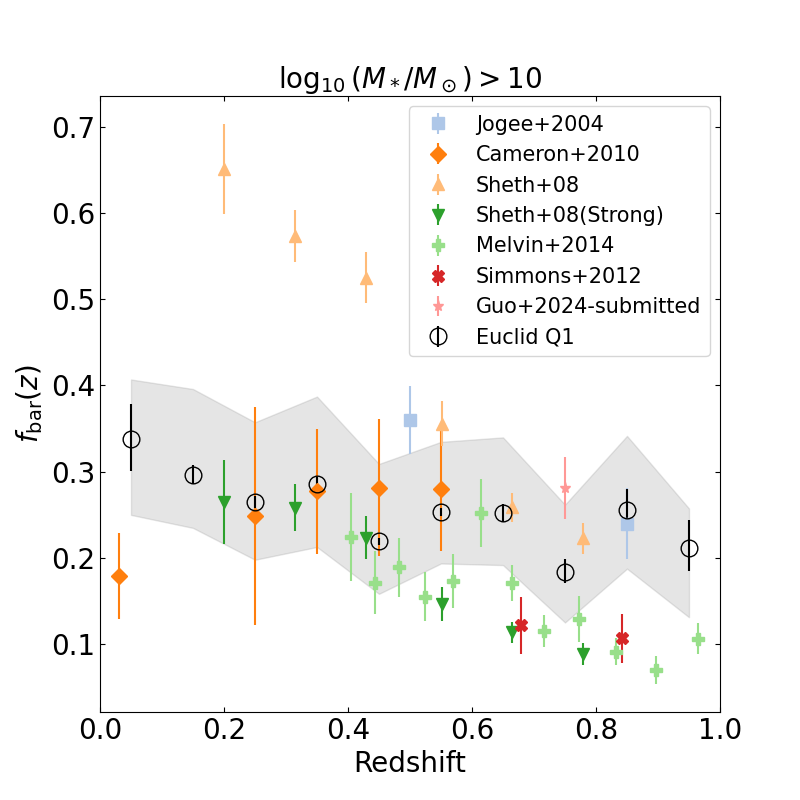}
 \caption{Bar fraction as a function of redshift. The large black circles show the measurement from Q1 presented in this work. The grey shaded region indicates the effect of changing the threshold for selecting barred galaxies between 0.4 and 0.6. Different colours and symbols indicate previously published results from different space-based surveys as labelled. The \Euclid measurements are in general agreement with previous works, but with significantly smaller statistical error bars.
 }
  \label{fig:bar_frac}
\end{figure}

Several previous publications have examined the evolution of the bar fraction over a similar redshift and stellar mass range, primarily using HST and JWST data~\citep{2004ApJ...615L.105J, 2008ApJ...675.1141S, 2010MNRAS.409..346C, 2014MNRAS.445.3466S, 2014MNRAS.438.2882M, 2024arXiv240906100G}. Performing a robust, apples to apples comparison with published results remains very difficult, given the variety of detection methods and sample selections. Some of these works are based on visual inspections by experts (e.g.,~\citealp{2008ApJ...675.1141S}), ellipse fitting (e.g.,~\citealp{2010MNRAS.409..346C, 2004ApJ...615L.105J}), and GZ classifications~(e.g.,~\citealp{2014MNRAS.438.2882M,2014MNRAS.445.3466S}).We note that although~\cite{2008ApJ...675.1141S} used two independent methods, we only report here the results from visual classifications, which should be more directly comparable with our measurements.  Nevertheless, a first-order comparison can still be informative to illustrate the scatter in the bar fractions resulting from different methodologies and to place the new Q1 data into a broader context. Figure~\ref{fig:bar_frac} thus compares the Q1 measurements from this work to a compilation of other results, and Table~\ref{tbl:barfrac_all} lists the number of galaxies in each redshift bin. For simplicity in the comparison, we include all galaxies more massive than $10^{10}\,M_\odot$ without additional mass constraints. This choice reflects that some works rely on luminosity-selected samples, each subject to different biases, making a fully homogenised stellar-mass selection unfeasible. However, all data shown in Fig.~\ref{fig:bar_frac} broadly target the massive or bright end of the galaxy population, approximately beyond the knee of the luminosity function. We emphasize that, for the \Euclid\ data, these selections cause a severe incompleteness effect (see discussion in Sect.~\ref{sec:biases}), and evolutionary trends need to be analysed for bins of stellar mass as done in Sect.~\ref{sec:results}.

The unprecedented sample size from \Euclid significantly reduces statistical
errors compared to HST or JWST studies. This showcases one of \Euclid's key strengths, combining high spatial resolution with a wide field of view. Indeed, the total number of barred galaxies in the Q1 survey ($7711$) already surpasses by more than an order of magnitude that in any previously published study beyond the local Universe.  As a result, the error budget of the \Euclid measurements is likely to be dominated by systematics, such as classification errors and incompleteness, as previously discussed.

Apart from the results of \citet{2008ApJ...675.1141S}, all measurements consistently yield bar fractions of $0.1{-}0.3$ within the explored redshift range, despite the varied methods and selection criteria, including recent JWST findings (Guo et al.\ 2024, submitted). This consistency reinforces the reliability of the \Euclid\ classifications used here and suggests that the bar fraction in massive galaxies out to $z \sim 1$ is well constrained to about $30\%$, providing a robust test for galaxy formation models. We stress that, although there is only one data point from JWST observations within the redshift range explored in this work, it is consistent within the uncertainties with the \Euclid measurements presented here. This is particularly important because, as described in Sect.~\ref{sec:bars}, the abundance of bars is known to decrease at shorter wavelengths because of outshining from young stellar populations and the effect of dust~\citep{2000ApJ...529...93K, 2018MNRAS.474.5372E}. The VIS filter being particularly wide~\citep{Q1-TP002}, these effects might be enhanced. The fact that the JWST NIR-based measurements provide similar values, however, suggests that wavelength variations within the redshift range explored do not severely affect our results. Although not explicitly shown in Fig.~\ref{fig:bar_frac}, previous studies~\citep{2010MNRAS.409..346C,2014MNRAS.438.2882M} and our own findings agree that more massive galaxies have systematically higher bar fractions (although see~\citealp{2012ApJ...761L...6M}). We expand on this mass dependence in Sect.~\ref{sec:sims}.

Nonetheless, the evolutionary trends reported by different authors show significant variation. This likely reflects the disparate sample selections and completeness limits, underlining the caution required when interpreting evolution in the bar fraction. For instance, \citet{2004ApJ...615L.105J} studied bars to $z \approx 1$ in the Galaxy Evolution from Morphologies and SEDs (GEMS) survey~\citep{2008ApJS..174..136C} and found a nearly constant bar fraction of $30\%\pm6\%$ based on various absolute-luminosity cuts, suggesting that dynamically cold discs were already established by $z \approx 1$. Similarly, \citet{2010MNRAS.409..346C} reported a flat trend. However, as shown in Fig.~\ref{fig:euclid_bar_frac_relative}, the bar-fraction evolution can appear artificially flattened if the sample is luminosity-limited, since brighter/more massive galaxies at higher redshifts intrinsically exhibit a larger bar fraction. This effect is also visible in our Q1 results (Fig.~\ref{fig:bar_frac}). Other works found a more pronounced decline in bar fraction with increasing redshift (e.g., \citealp{2008ApJ...675.1141S,2014MNRAS.445.3466S, 2014MNRAS.438.2882M}). \citet{2014MNRAS.445.3466S} employed a redshift-dependent luminosity cut that may counteract the mass dependence, while \citet{2014MNRAS.438.2882M} used a stellar-mass selection similar to ours but with deeper COSMOS data, potentially explaining their stronger evolution at higher redshifts. Notably, \citet{2008ApJ...675.1141S} reported both a higher bar fraction and a steeper redshift dependence, likely due to a combination of selection effects and a classification scheme that includes both strong and weak bars. When only strong bars are considered, their measurements align more closely with ours, implying that our GZ-based classifications primarily capture strong bars.

\begin{table}
\centering
  \caption{Same as Table~\ref{tbl:nbars} but for all galaxies more massive than $10^{10}M_\odot$. The last row shows the total amount of barred and featured galaxies in the sample analysed in this work. We emphasize that the bar fraction is severely affected by incompleteness -- see text for details. }
  \begin{tabular}{lcrrcl}
  \hline
  \hline
     & $z_\textrm{min}$--$z_\textrm{max}$ & $N_\textrm{featured}$ & $N_\textrm{bar}$ & $f_\textrm{bar}$ & \\ 
    \hline
    \noalign{\vskip 4pt}
    \multicolumn{6}{c}{ ($\logten(M_*/M_\odot) > 10$)} \\ 
    \hline
    \noalign{\vskip 4pt}
     & 0.00--0.10 & 145 & 49 & 0.338 & \\ 
     & 0.10--0.20 & 1676 & 497 & 0.297 & \\ 
     & 0.20--0.30 & 3559 & 943 & 0.265 & \\ 
     & 0.30--0.40 & 3599 & 1030 & 0.286 & \\ 
     & 0.40--0.50 & 10\,129 & 2225 & 0.220 & \\ 
     & 0.50--0.60 & 8876 & 2249 & 0.253 & \\ 
     & 0.60--0.70 & 1771 & 446 & 0.252 & \\ 
     & 0.70--0.80 & 795 & 146 & 0.184 & \\ 
     & 0.80--0.90 & 338 & 86 & 0.254 & \\ 
     & 0.90--1.00 & 191 & 40 & 0.209 & \\ 
     \hline
     \noalign{\vskip 4pt}
      &\textbf{ 0.00--1.00} & \textbf{27\,480} & \textbf{7711} & \textbf{0.280} & \\ 
    \hline
  \end{tabular}  \label{tbl:barfrac_all}
\end{table}

\subsection{Comparison with cosmological simulations}
\label{sec:sims}

 Comparing observed bar properties with simulations helps identify key processes
driving disc assembly over time. This is the focus of this subsection. Guided by the discussion in Sect.~\ref{sec:lit}, we restrict our comparison to the two main robust findings of this work: (1) the average bar fraction over $z=0{-}1$, and (2) its stellar mass dependence. We consider two recent, state-of-the-art simulations, \texttt{TNG50} and \texttt{Auriga}. It is important to note that comparing observations and simulations is not free from biases. In particular, as discussed before, observational data are subject to various selection effects not present in simulations. A fully robust, `apples-to-apples'
 comparison would require a forward modelling of the simulation outputs into the observational plane, which is beyond the scope of this work (e.g.,~\citealp{2021MNRAS.501.4359Z}). For example, the bar fraction in observations is computed over a sample of featured galaxies (Sect.~\ref{sec:bars}), while in simulations bars are quantified in disc galaxies selected based on their dynamics. Although the featured label serves as a proxy for disc galaxies, it does not imply a perfect correspondence, which can accentuate certain discrepancies.

The \texttt{TNG50} simulation is part of the IllustrisTNG project, a suite of cosmological simulations aimed at exploring galaxy formation and evolution~\citep{2018MNRAS.475..648P}. These simulations employ the \texttt{AREPO} moving-mesh code~\citep{2010MNRAS.401..791S}, which accounts for gravitational interactions and incorporates sub-grid models to capture baryonic processes, building upon earlier work from the Illustris project~\citep{2014MNRAS.445..175G,2014Natur.509..177V}. \texttt{TNG50} has the smallest volume (50 comoving Mpc) of the suite, but offers higher resolution ($ 8.5\times10^4\,M_\odot$), making it suitable for probing the internal structure of galaxies. For the comparison presented here, we use the results of~\cite{2022MNRAS.512.5339R} and~\cite{2024MNRAS.529..979L}, who analysed the bar fraction and discussed bar formation in \texttt{TNG50}.The sample of ~\cite{2022MNRAS.512.5339R}is comprised of a complete sample of galaxies more massive than $10^{10}M_\odot$ with a disc-to-total ratio ($D/T$) larger than 0.5. 

The \texttt{Auriga} simulation is another set of cosmological magneto hydrodynamical zoom-in simulations of individual halos spanning $M_{\rm 200}\in[0.5,2.0]\times 10^{12} M_\odot$ at $z=0$~\citep{2017MNRAS.467..179G}. They also use the AREPO code, but with a slightly different galaxy-formation model (see~\citealp{2013MNRAS.436.3031V,2014MNRAS.437.1750M,2017MNRAS.467..179G} for details), which includes cooling, background UV fields for reionisation, subgrid prescriptions for star formation, stellar evolution and feedback, magnetic fields, and black hole seeding, accretion, and feedback. The stellar and gas mass resolution is $5\times10^4\,M_\odot$. We compare with \citet{2024arXiv240609453F}, who studied the properties of barred galaxies in \texttt{Auriga}. We stress that since these simulations are zoom-in, the measurements reported correspond to a representative instead of a complete sample of galaxies. All galaxies have stellar masses larger than $10^{10}M_\odot$ at $z=0$ and are for the vast majority disc dominated ($D/T>0.5$). More details can be found in~\cite{2017MNRAS.467..179G}.

\begin{figure}
\includegraphics[width=\columnwidth]{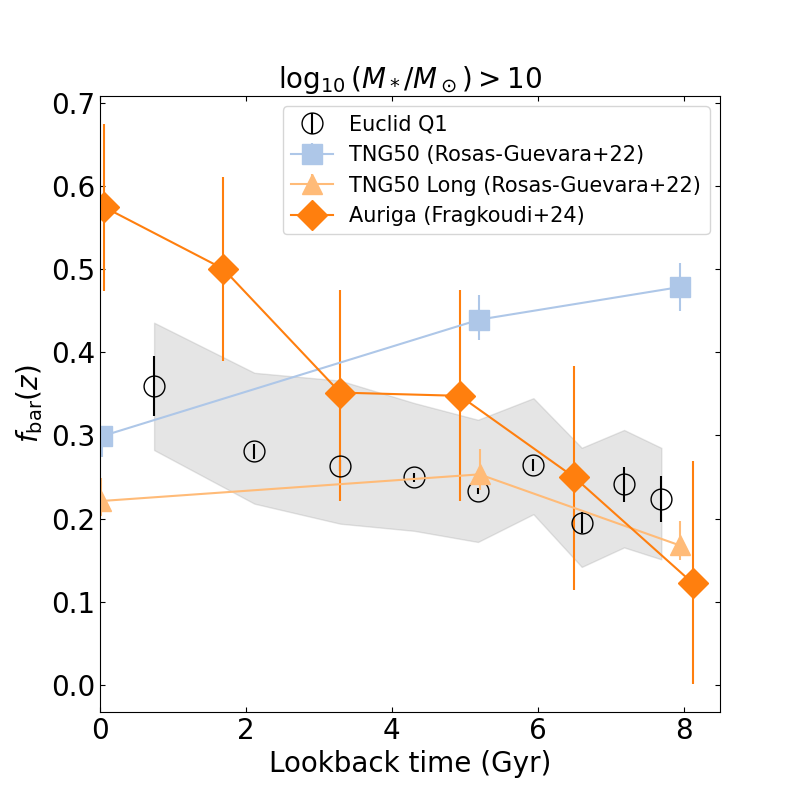}

 \caption{Comparison of the observed bar fraction in our \Euclid sample (large empty circles) with cosmological simulations. The cyan squares and pink triangles show the results from the \texttt{TNG50} simulation when all bars and only long bars are included respectively. The orange diamond shows the \texttt{Auriga} simulation. The grey shaded region indicates the effect of varying the probability threshold for bar selection between 0.4 and 0.6. The mean bar fraction is globally well reproduced by the simulations. 
 }
  \label{fig:bar_frac_sims}
\end{figure}

Figure~\ref{fig:bar_frac_sims} shows the evolution of the bar fraction for the complete Q1 sample with $M_* > 10^{10}\,M_\odot$ alongside the \texttt{Auriga} and \texttt{TNG50} predictions, applying the same stellar-mass selection to the simulations. Hence, these bar fractions do not trace the true progenitor evolution. Interestingly, both simulations predict a global bar fraction consistent with the observational values of about $0.2{-}0.4$. The redshift trends differ, but as stressed above, one cannot draw strong conclusions without matching selection effects in the simulations and observations. Nevertheless, both simulations seem to produce a lower fraction at low redshift, discussed further below. For \texttt{TNG50}, we also show the bar fraction when all bars are included, even very small ones that may be difficult to detect in the observations. In this case, the bar fraction becomes significantly larger than the observational estimates, emphasising the complexity of simulation--observation comparisons and the importance of carefully modelling selection biases. 

\begin{figure*}
\includegraphics[width=\columnwidth]{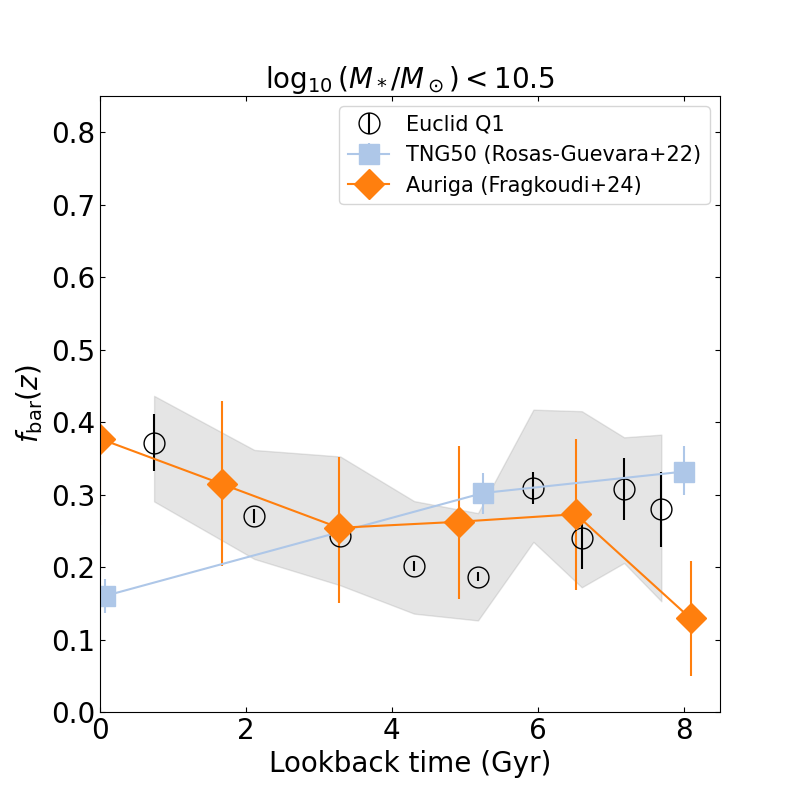}
\includegraphics[width=\columnwidth]{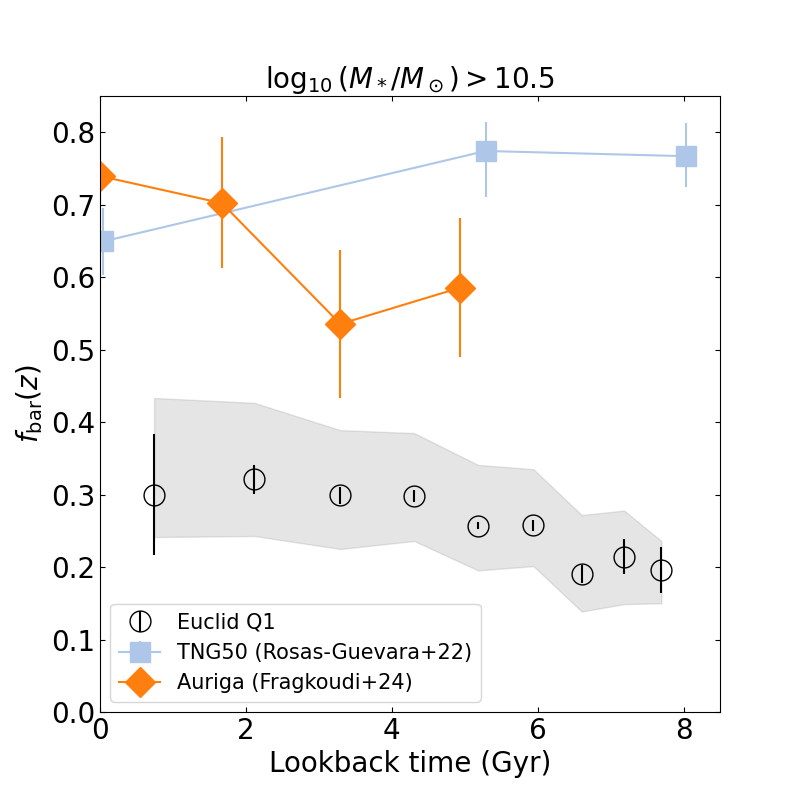}

 \caption{Same as Fig.~\ref{fig:bar_frac_sims}, but dividing galaxies in stellar mass bins. The left panel shows galaxies with stellar masses between $10^{10}$ and $10^{10.5}$ $M_\odot$. The right panel shows galaxies more massive than $10^{10.5}$ $M_\odot$. Simulations tend to over predict the bar fraction at the high mass end.
 }
  \label{fig:bar_frac_sims_massbins}
\end{figure*}

Another key result from this work is the stellar mass dependence of the bar fraction. Figure~\ref{fig:bar_frac_sims_massbins} repeats the comparison with simulations, now split into two stellar mass bins. While \texttt{TNG50} uses $\log_{10}(M_*/M_\odot)=10.5$ and \texttt{Auriga} adopts $10.7$, we find the overall trends remain similar regardless of the exact division. Both simulations predict a clear mass dependence in bar fraction, with more massive galaxies hosting more bars. However, the effect is more pronounced in the simulations, where approximately $70\%$ of massive galaxies are barred, compared to only $40\%$ in the observations. 

Early idealised simulations suggested that high gas fractions can inhibit bar formation (e.g., \citealp{1986MNRAS.221..213A,2003MNRAS.341.1179A,2010ApJ...719.1470V}), but more recent cosmological simulations such as \texttt{TNG50} and \texttt{Auriga}, indicate that the ratio of baryonic to dark matter is a primary factor in regulating bar formation. \citet{2024MNRAS.529..979L} observed that unbarred galaxies consistently have lower
central baryonic-to-dark matter ratios than barred galaxies. Similarly, \citet{2024arXiv240609453F} found no significant difference in gas fractions between
barred and unbarred galaxies at a fixed stellar mass. Interestingly, the latter study also notes that baryon-dominated galaxies without bars often have richer merger histories.  \citet{2022MNRAS.512..160R}, using the \texttt{NEWHORIZON} simulation, found that excessive
dark matter or large bulges could inhibit bar formation. Therefore, the higher bar fractions in high-mass galaxies predicted by simulations compared to observations may reflect overly efficient central star formation, which boosts the baryonic-to-dark-matter ratio and thus favours bar formation. Alternatively, simulated galaxies may experience fewer mergers, resulting in lower ex-situ fractions. However, recent work on local Universe ex-situ stellar mass fractions~\citep{2024NatAs...8.1310A} suggests that the integrated merger rate is relatively well reproduced by state-of-the-art simulations.

\section{Summary and conclusions}\label{sec:summary}

 We have investigated the abundance of stellar bars in massive disc galaxies (\(M_* \gtrsim 10^{10}\,M_\odot\)) up to \(z \approx 1\) using data from the Q1 release over an area of 63.1\,deg$^2$. By applying a deep-learning model trained on citizen science visual labels, we identified barred galaxies in a magnitude-selected sample (\(\IE < 20.5\)). Our main findings can be summarised as follows.

\begin{itemize}
    \item 
    We have identified \(7711\) barred galaxies between \(z=0\) and \(z=1\), exceeding, by an order of magnitude, the samples from previous work over a similar redshift range. This highlights \Euclid's unique capability to resolve internal galaxy structures across a wide sky area.

    \item 
    The mean bar fraction of 0.2--0.4 agrees well with estimates from HST-based 
    surveys, indicating that \Euclid\ can robustly reproduce morphological measures over large samples.  In the era of very large surveys such as \Euclid, a proper quantification of systematic effects such as classification errors becomes extremely important. 

    \item 
    At a fixed redshift, massive systems exhibit a higher bar fraction than lower-mass galaxies, and the decrease in bar fraction with redshift is more pronounced for lower-mass systems. This suggests an earlier formation and assembly of discs in massive galaxies.

    \item 
    While cosmological simulations match the overall bar fraction, they overpredict it for the most massive galaxies. This discrepancy suggests that the models may overestimate the efficiency of central stellar mass growth.
\end{itemize}

Overall, these results illustrate the effectiveness of \Euclid's combination of spatial resolution and wide-area coverage in probing the internal structure of disc galaxies. Future work includes incorporating finer mass bins, additional morphological indicators, more detailed comparisons with simulations, and correlation with environmental indicators that have not been addressed in this first work. The \Euclid data will indeed enable a unique quantification of large-scale structure~\citep{Q1-SP028} enabling a precise dissection of the role of environment in shaping galaxy structure.

\begin{acknowledgements} 
\AckQone \AckEC MHC acknowledges support from the State Research
Agency (AEIMCINN) of the Spanish Ministry of Science and Innovation under the grant "BASALT" with reference PID2021-126838NBI00. Co-funded by the European Union (Widening Participation, ExGal-Twin, GA 101158446 and MSCA Doctoral Network EDUCADO, GA 101119830).
\end{acknowledgements} 
\bibliographystyle{aa}
\bibliography{Euclid, Q1, biblio}

\label{LastPage}

\end{document}